\documentclass[aps,prl,twocolumn,superscriptaddress,floatfix,nofootinbib,showpacs,amsmath,amssymb,amsart]{revtex4-1}
\input{bibliography.def}
\usepackage{array}
\usepackage{siunitx}
\newcolumntype{L}[1]{>{\raggedright\let\newline\\\arraybackslash\hspace{0pt}}p{#1}}
\newcolumntype{C}[1]{>{\centering\arraybackslash}p{#1}}
\usepackage{microtype}
\usepackage{lineno}
\usepackage{comment}
\usepackage{nameref}
\makeatletter
\newcommand\setcurrentname[1]{\def\@currentlabelname{#1}}
\makeatother

\begin{document}
\setcounter{secnumdepth}{5}
\title{Dark matter search with the \DEAP\ detector using the profile likelihood ratio method}
\newcommand{\Alberta}{Department of Physics, University of Alberta, Edmonton, Alberta, T6G 2R3, Canada}
\newcommand{\AstroCeNT}{AstroCeNT, Nicolaus Copernicus Astronomical Center, Polish Academy of Sciences, Rektorska 4, 00-614 Warsaw, Poland}
\newcommand{\BHSU}{School of Natural Sciences, Black Hills State University, Spearfish, SD 57799, USA}
\newcommand{\Cagliari}{Physics Department, Universit\`a degli Studi di Cagliari, Cagliari 09042, Italy}
\newcommand{\UCR}{Department of Physics and Astronomy, University of California, Riverside, CA 92521, USA}
\newcommand{\CNL}{Canadian Nuclear Laboratories, Chalk River, Ontario, K0J 1J0, Canada}
\newcommand{\Carleton}{Department of Physics, Carleton University, Ottawa, Ontario, K1S 5B6, Canada}
\newcommand{\CIEMAT}{Centro de Investigaciones Energ\'eticas, Medioambientales y Tecnol\'ogicas, Madrid 28040, Spain}
\newcommand{\Houston}{Department of Physics, University of Houston, Houston, TX 77204, USA}
\newcommand{\INFNCagliari}{INFN Cagliari, Cagliari 09042, Italy}
\newcommand{\INFNNapoli}{INFN Napoli, Napoli 80126, Italy}
\newcommand{\Mainz}{Institut f\"ur Kernphysik, Johannes Gutenberg-Universit\"at Mainz, 55128 Mainz, Germany}
\newcommand{\LNGS}{INFN Laboratori Nazionali del Gran Sasso, Assergi (AQ) 67100, Italy}
\newcommand{\LU}{School of Natural Sciences, Laurentian University, Sudbury, Ontario, P3E 2C6, Canada}
\newcommand{\LBNL}{Nuclear Science Division, Lawrence Berkeley National Laboratory, Berkeley, CA 94720, USA}
\newcommand{\UNAM}{Instituto de F\'isica, Universidad Nacional Aut\'onoma de M\'exico, A.\,P.~20-364, Ciudad de M\'exico~01000, Mexico}
\newcommand{\Napoli}{Physics Department, Universit\`a degli Studi ``Federico II'' di Napoli, Napoli 80126, Italy}
\newcommand{\Capodimonte}{Astronomical Observatory of Capodimonte, Salita Moiariello 16, I-80131 Napoli, Italy}
\newcommand{\MEPhI}{National Research Nuclear University MEPhI, Moscow 115409, Russia}
\newcommand{\Oxford}{Department of Physics, University of Oxford, Oxford, OX1 3PU, United Kingdom}
\newcommand{\IIPA}{International Institute for Particle Astrophysics, Polish Academy of Sciences,  Bartycka 18, 00-716 Warsaw, Poland}
\newcommand{\Princeton}{Physics Department, Princeton University, Princeton, NJ 08544, USA}
\newcommand{\Queens}{Department of Physics, Engineering Physics, and Astronomy, Queen's University, Kingston, Ontario, K7L 3N6, Canada}
\newcommand{\RHUL}{Royal Holloway University London, Egham Hill, Egham, Surrey TW20 0EX, United Kingdom}
\newcommand{\RAL}{Rutherford Appleton Laboratory, Harwell Oxford, Didcot OX11 0QX, United Kingdom}
\newcommand{\SL}{SNOLAB, Lively, Ontario, P3Y 1M3, Canada}
\newcommand{\Sussex}{University of Sussex, Sussex House, Brighton, East Sussex BN1 9RH, United Kingdom}
\newcommand{\TRIUMF}{TRIUMF, Vancouver, British Columbia, V6T 2A3, Canada}
\newcommand{\TUD}{Institut für Kern und Teilchenphysik, Technische Universität Dresden, 01069 Dresden, Germany}
\newcommand{\TUM}{Department of Physics, Technische Universit\"at M\"unchen, 80333 Munich, Germany}
\newcommand{\MI}{Arthur B. McDonald Canadian Astroparticle Physics Research Institute, Queen's University, Kingston, ON, K7L 3N6, Canada}

\affiliation{\Alberta}
\affiliation{\AstroCeNT}
\affiliation{\BHSU}
\affiliation{\Cagliari}
\affiliation{\UCR}
\affiliation{\CNL}
\affiliation{\Carleton}
\affiliation{\CIEMAT}
\affiliation{\Houston}
\affiliation{\INFNCagliari}
\affiliation{\INFNNapoli}
\affiliation{\Mainz}
\affiliation{\LU}
\affiliation{\UNAM}
\affiliation{\Capodimonte}
\affiliation{\Napoli}
\affiliation{\MEPhI}
\affiliation{\Oxford}
\affiliation{\IIPA}
\affiliation{\Queens}
\affiliation{\RHUL}
\affiliation{\RAL}
\affiliation{\SL}
\affiliation{\TRIUMF}
\affiliation{\TUD}
\affiliation{\TUM}
\affiliation{\MI}

\author{P.~Adhikari}\affiliation{\Carleton}
\author{R.~Ajaj}\affiliation{\Carleton}\affiliation{\MI}
\author{M.~Alp\'izar-Venegas}\affiliation{\UNAM}
\author{P.-A.~Amaudruz}\affiliation{\TRIUMF}
\author{J.~Anstey}\affiliation{\Carleton}\affiliation{\MI}
\author{D.\,J.~Auty}\affiliation{\Alberta}
\author{M.~Baldwin}\affiliation{\RAL}
\author{M.~Batygov}\affiliation{\LU}
\author{B.~Beltran}\affiliation{\Alberta}
\author{A.~Bigentini}\affiliation{\Carleton}\affiliation{\MI}
% ...

\collaboration{DEAP Collaboration}
\altaffiliation{deap-papers@snolab.ca}

\date{\today}

\begin{abstract}
\begin{center}
\end{center}
We present here a search for WIMP dark matter using \PaperThreeLiveTimeNum\ live-days of data collected with \larmass\ of liquid argon (1266~kg fiducial) by the \DEAP\ detector at SNOLAB, using the \ProfileLikelihoodRatio\ method.
The likelihood model is based on three parameters: estimated energy, pulse-shape discrimination parameter, and reconstructed position within the detector. 
Using this method, the expected signal sensitivity of \DEAP\ benefits from an increased fiducial volume and improved event selection acceptance.
Alpha-decays from a small number of dust particulates circulating within the liquid argon target are the dominant source of background events and limit the sensitivity of this search.
This result provides improved exclusion upper limits on the WIMP-nucleon spin-independent cross section on liquid argon for WIMP masses between \SI{20}{GeV}/$c^{2}$ and \SI{100}{GeV}/$c^{2}$.
At \SI{100}{GeV}/$c^{2}$ the observed limit is
\PaperThreeWIMPLimitOneHundredGeV\
at 90\% confidence level.
\end{abstract}

\maketitle

\section*{Introduction}\setcurrentname{Introduction}\label{sec:intro}
There is compelling evidence for the existence of a non-luminous type of matter in our Universe, known as dark matter, that is outside the Standard Model of particle physics. Astrophysical and cosmological observations have established that dark matter comprises 27\% of the total energy density of the Universe, and is approximately five times more abundant than the ordinary matter component~\cite{planck_collaboration_planck_2018}. Despite this, dark matter has yet to be experimentally observed. A variety of dark matter candidates with the required properties to reproduce the observed dark matter relic density have been theorized; one such candidate is Weakly Interacting Massive Particles (WIMPs). Elastic scattering of WIMP particles with Standard Model particles, such as atomic nuclei, have the potential to be observed in ``direct detection'' experiments, designed to search for low energy $\mathcal{O}(1-100\ \mathrm{keV})$ nuclear recoils (NRs) induced by WIMP interactions.

The \DEAP\ dark matter experiment~\cite{amaudruz_design_2017} at SNOLAB is a direct detection experiment using liquid argon (LAr) as its target medium to search for WIMP dark matter.
Liquid argon is an excellent choice of target medium as it is easy to purify, has high scintillation light yield, is transparent to its own scintillation light, and features outstanding pulseshape discrimination (PSD)~\cite{DEAP_PSD} between NRs and electronic recoils (ERs).
Based on 231 live-days of data, \DEAP\ previously placed upper limits at 90\% confidence level (CL) on the WIMP-nucleon spin-independent cross section $\sigma_{\chi}$ as a function of the WIMP mass $M_{\chi}$, with a limit of \PaperTwoWIMPLimitOneHundredGeV\ at 100~GeV/$c^{2}$ and of \PaperTwoWIMPLimitOneTeV\ at 1~TeV/$c^{2}$~\cite{deapdarkmatter2019}.

We present new WIMP dark matter search results using a multi-dimensional \ProfileLikelihoodRatio\ (PLR) statistical method on the \PaperThreeLiveTimeNum\ live-days of \DEAP\ data collected from November 2016 to March 2020. 
From January 2018, 80\% of the physics data were designated ``blind'' for the WIMP search, and were only unveiled in November 2025 once the methods of the search were frozen and internally reviewed by the collaboration: the exposure corresponding to this part of the dataset is \PaperThreeBlindLiveTimeNum~live-days, i.e. \PaperThreeOpenLiveTimeNum~live-days constituted the ``open'' dataset.

The \PLR\ approach is popular within the dark matter direct detection community 
as it
allows an experiment to expand the signal search region and loosen background rejection cuts in order to increase the signal acceptance. 
The main advantage of the PLR approach is that it makes use of observable distributions to reduce the impact of backgrounds in the region of interest (ROI), in order to maximize an experiment's sensitivity. 

With respect to the previous \DEAP\ result~\cite{deapdarkmatter2019}, the signal sensitivity benefits from a longer
exposure, from a wider ROI in the PSD variable \FPrompt, from removing the previously-used event selection cut on the observable \PIFGAr\ (as described under Methods), as well as from a significant increase
in the fiducial volume defined with position reconstruction.

This work introduces dust $\alpha$-decays from a low level of particulate contamination in the LAr as a new component of the \DEAP\ background model.
This background component, that was under consideration but not formally part of our model in Ref.~\cite{deapdarkmatter2019}, is 
estimated using a dedicated dust control region (CR) and indeed found to be the dominant background in the updated WIMP ROI, limiting the sensitivity that can be achieved with this dataset.

Further analysis on the \DEAP\ data is being carried out, using machine-learning techniques
to improve event reconstruction including the discrimination against $\alpha$-decay background events.

Hardware upgrades to the \DEAP\ detector have been installed, the main objectives of which being to mitigate $\alpha$-decay backgrounds from the detector neck and from dust particulates.  As of summer 2025, data taking with LAr has started in the upgraded configuration.

\section*{Results and Discussion}\setcurrentname{Results and Discussion}\label{sec:results}

\subsection*{Event Selection, ROI Definition, and Signal Acceptance}\setcurrentname{Event Selection, ROI Definition and Signal Acceptance}\label{subsec:resultsROI}

In this search, the data are subjected to event selection cuts, grouped into categories as shown in Table~\ref{T:Cuts}.
(1)~Data quality cuts are standard instrumental data-cleaning cuts that are applied to any physics data acquired by the detector. 
(2)~Pile-up cuts are designed to remove events that are suspected to be coincidence events, where there is evidence of two or more individual physics signals in one event window. 
(3) Events where significant light is seen in the outer detector's muon veto \PMTs\ or in the neck veto \PMTs\ are removed.
(4)~Fiducial cuts are used to remove events that reconstruct near the edge of the detector, such as surface $\alpha$-decays and radiogenic neutron background events. 
(5)~Background rejection cuts are used to further remove backgrounds; these cuts are particularly effective against shadowed $\alpha$-decays.
(6)~A~WIMP search ROI is defined in the two-dimensional parameter space \FPrompt\ versus the number of detected photoelectrons (PE).

The event selection cuts have been optimized using the model detailed in Section: \Nameref{sec:models}, using a combination of theoretical models, data observed in CRs and sidebands, simulated data, and calibration data, without regard to the number of events actually observed in the WIMP ROI.

The same event selection cuts as used in Ref.~\cite{deapdarkmatter2019} are applied to define the WIMP ROI in this analysis, with the following exceptions:
\begin{itemize}
\item In the definition of the fiducial volume, the value of \R\ is increased from \SI{630}{mm} to \SI{720}{mm}. This results in a net increase of 54\% in fiducial mass~\cite{delgobbo_2021,deapPosRec} to 1266 kg, taking into account all fiducial cuts.
\item The ROI ceiling in \FPrompt\ is higher, resulting in a 36\% increase in signal acceptance at \mbox{$M_\chi = 100$ GeV/$c^{2}$}.
\item The search threshold in PE is slightly reduced, from \SI{94}{PE} to \SI{90}{PE}. This results in a small 3\% increase in signal acceptance at \mbox{$M_\chi = 100$ GeV/$c^{2}$}.
\item The upper boundary of the ROI in PE is increased from 200~PE to 300~PE.
This results in a 22\% increase in signal acceptance at \mbox{$M_\chi = 100$ GeV/$c^{2}$} (the increase is larger at higher $M_\chi$) and allows for a better constraint on $\alpha$-decay backgrounds in the PLR fit.
\item The cut on \PIFGAr\ is no longer applied. This results in a 76\% relative increase in signal acceptance integrated over~90--300~PE.
\end{itemize}

These improvements in signal acceptance are made possible by the adoption of the PLR approach, where accounting for the background model in the final fit allows, in principle, for partial recovery of the sensitivity loss due to the presence of backgrounds in the ROI.
The list of event selection cuts applied to define this updated ROI and the CR for dust $\alpha$-decays is given in Table~\ref{T:Cuts}.

\begin{table}[ht]
\begin{center}
\caption{Event selection cuts used to define the dust CR and the WIMP ROI in this analysis.}
\label{T:Cuts}
\begin{tabular}{lcc}
\hline \hline
\textbf{Cut name}     &    \textbf{Dust \CR}     &    \textbf{\WIMP\ \ROI}   \\  \hline
Data quality     &  $\checkmark$   &  $\checkmark$    \\
Pile-up cuts    &    $\checkmark$   &  $\checkmark$     \\
Muon veto    &    --   &  $\checkmark$     \\
Neck veto    &    --   &  $\checkmark$     \\
\hline
\textit{Fiducial cuts:} \\
Fraction of \PE\ observed in   &  $< 0.4$        &    $< 0.4$     \\
~~~~a single \PMT\  &                &        \\
Fraction of \PE\ observed in   &  $< 0.04$        &    $< 0.04$     \\
~~~~top 2 rows of \PMTs\  &                &        \\
Fraction of \PE\ observed in   &  $< 0.1$        &    $< 0.1$     \\
~~~~bottom 3 rows of \PMTs\  &                &        \\
$z_{\rm rec}$ above origin    &    $< \SI{550}{mm}$   &  $< \SI{550}{mm}$    \\
\R\ from origin   &    $< \SI{720}{mm}$     &   $< \SI{720}{mm} $    \\
\hline
\textit{Against shadowed $\alpha$-decays:} \\
Position reconstruction   &    --   &  $\checkmark$     \\
~~~~consistency      &                &        \\
\PIFGAr  &   $> 2$       &      --     \\
\hline
\textit{Energy and PSD:} \\
\PE\ range  &  500--20000  &  90--300  \\
\FPrompt    & $> 0.5$   &  Contour  \\
\hline \hline
\end{tabular}
\end{center}
\end{table}

\subsection*{Results}\setcurrentname{Results}\label{subsec:results_final}

\begin{figure*}[t!b]
 \centering
 \includegraphics[width=0.49\linewidth]{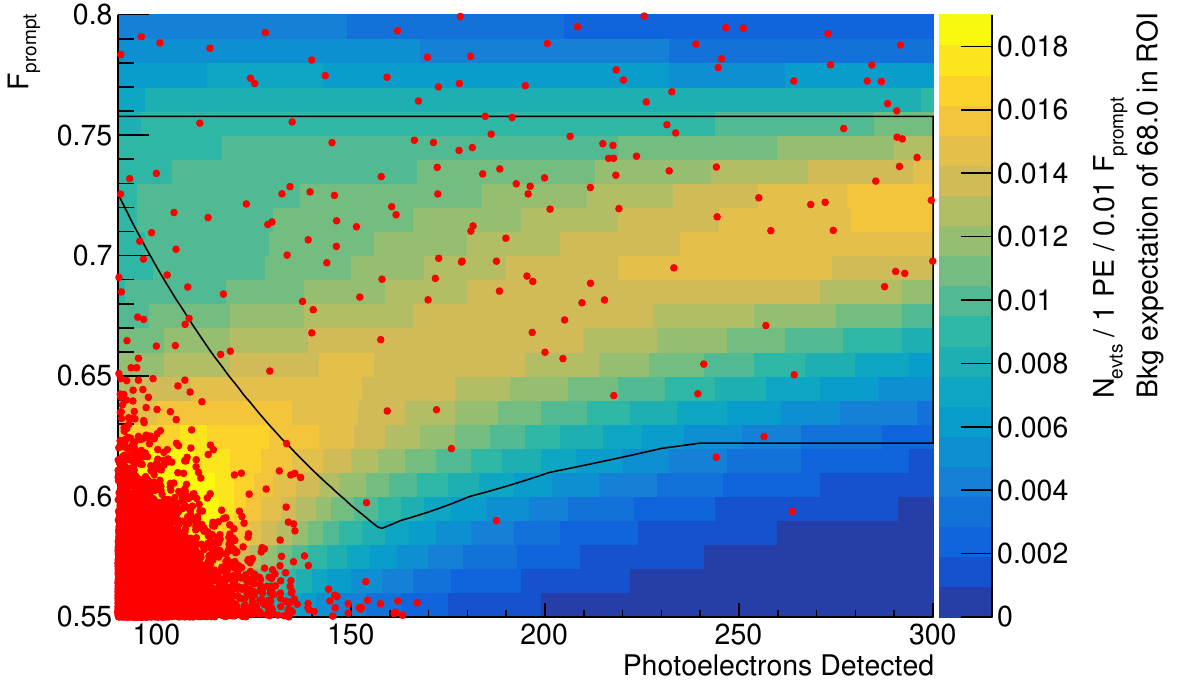}
 \includegraphics[width=0.49\linewidth]{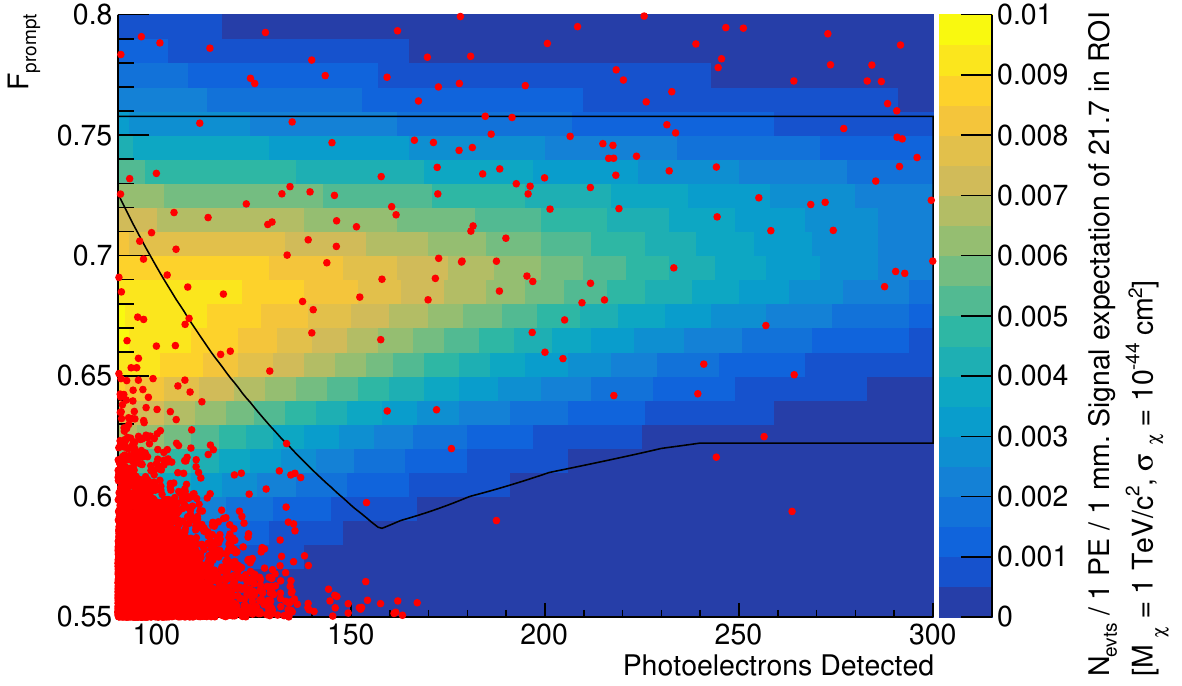}
 \includegraphics[width=0.49\linewidth]{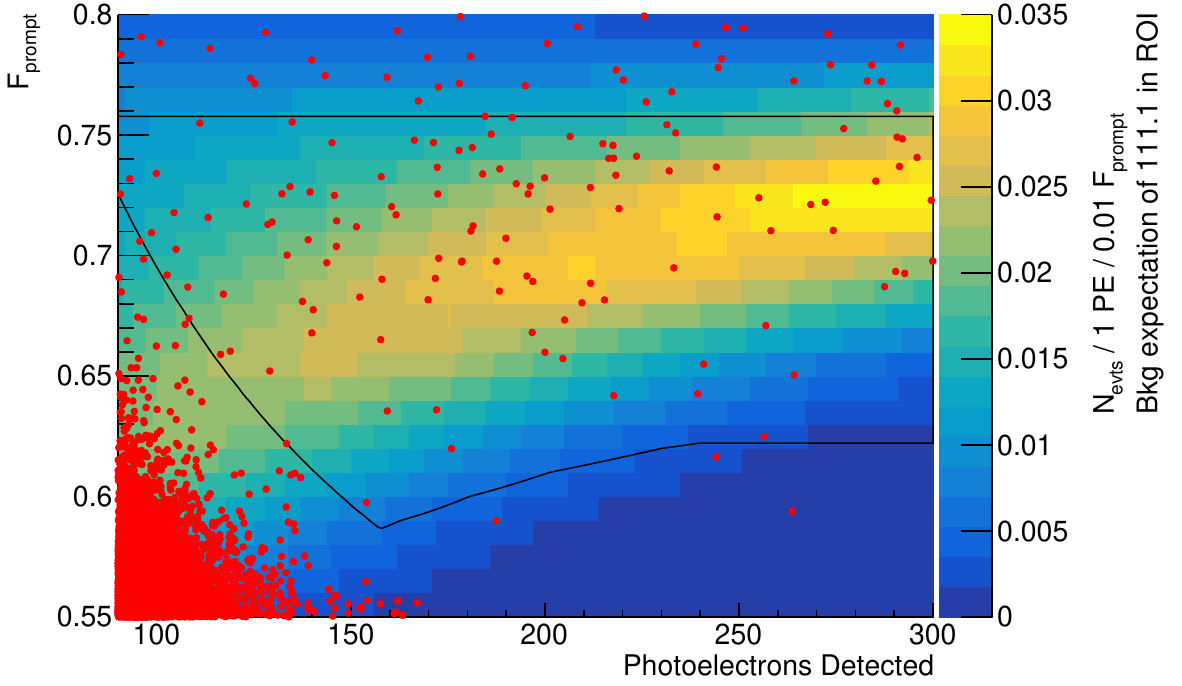}
 \includegraphics[width=0.49\linewidth]{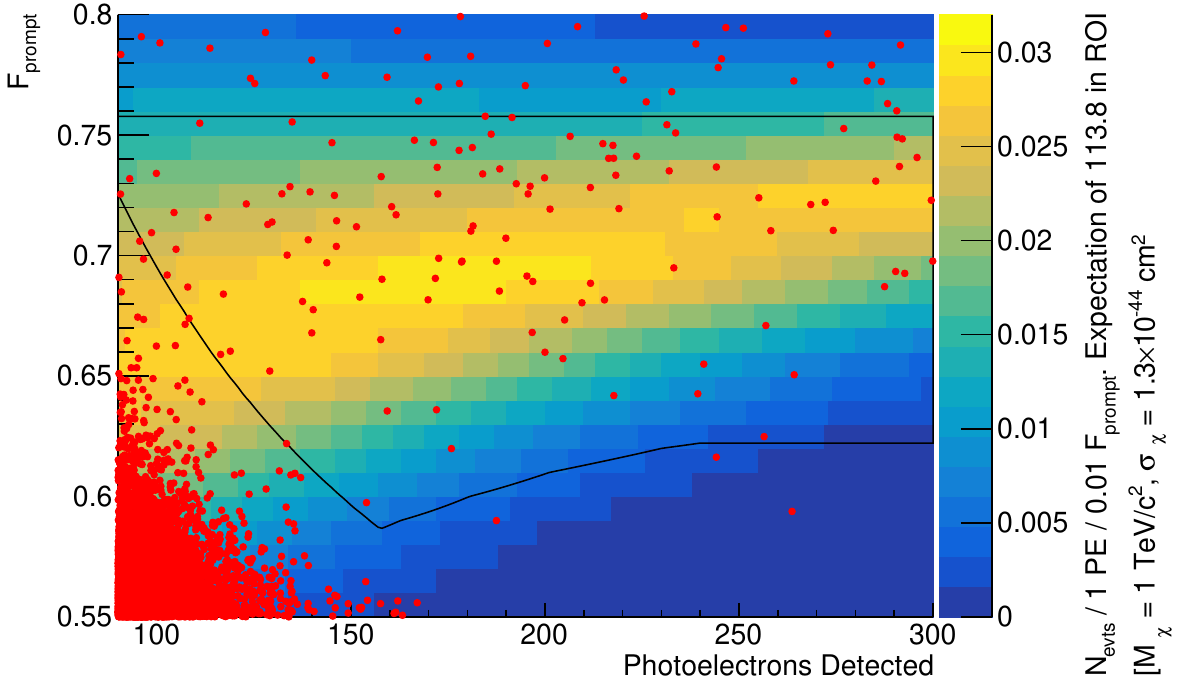}
 \caption{\ROI\ contour in \FPrompt\ versus \PE, illustrating the events passing all event selection criteria (red points) superimposed on:
 (top-left) the prior background model,
 (top-right) the \WIMP\ model assuming a signal hypothesis of \mbox{$M_{\chi}$ = 1 TeV/$c^{2}$},
 \mbox{$\sigma_{\chi}$ = 10$^{-44}$ cm$^{2}$},
 (bottom-left) the best-fit background model under the background-only hypothesis, and
 (bottom-right) the background+signal best-fit model with \mbox{$M_{\chi}$ = 1 TeV/$c^{2}$}.
 Where applicable, the background model shown is the sum of radiogenic neutrons, surface $\alpha$-decays, shadowed $\alpha$-decays, and dust $\alpha$-decays.
 $^{39}$Ar $\beta$-decay events are seen outside the ROI to the bottom-left of each plot; this background is negligible within the ROI.
 }
 \label{F:ROIwithData}
\end{figure*}

After applying all event selection cuts, there are 115 events observed inside the \ROI: 40 in the open dataset and 75 in the blind dataset.
This is the data sample upon which the \PLR\ analysis is performed. 
These events are shown in \PE\/-\FPrompt\ space in Figure~\ref{F:ROIwithData} superimposed on the summed background model.
Figure~\ref{F:ROIwithData} also shows the \ROI\ drawn on top of the \WIMP\ model, assuming a signal hypothesis of \mbox{$M_{\chi}$ = 1 TeV/$c^{2}$}, \mbox{$\sigma_{\chi}$ = 10$^{-44}$ cm$^{2}$}, the best-fit background model under the background-only hypothesis, and the best-fit model assuming a signal hypothesis with \mbox{$M_{\chi}$ = 1 TeV/$c^{2}$}.
In the latter case, the best-fit \mbox{$\sigma_{\chi} = 1.3 \times 10^{-44}$ cm$^2$}.

Table~\ref{T:BackgroundPredictionsPost} compares the number of expected events from each source for the prior, for the posterior which corresponds to the best fit to the data assuming the background-only hypothesis, and for the best fit to the data under a signal hypothesis of \mbox{$M_{\chi}$ = 1 TeV/$c^{2}$}.  After the background-only fit, the total background in the \ROI\ is estimated to be \mbox{$111.1 ^{+ 16.3}_{- 15.2}$} events, consistent with the observed number of events.

\begin{table}[ht]
\begin{center}
\caption{Summary table of the expected number of \ROI\ events from each source in the prior background model, the posterior model in the background-only hypothesis, and the posterior model in the signal+background hypothesis with \mbox{$M_{\chi} = 1$ TeV/$c^{2}$}, \mbox{$\sigma_{\chi} = 1.3 \times 10^{-44}$ cm$^2$}.
Shown are the systematic uncertainties.
In data, 115~events are observed in the \ROI.
}
\label{T:BackgroundPredictionsPost}
\begin{tabular}{lccc}
\hline \hline
\textbf{Source}     &    \textbf{Prior}     &    \textbf{Best-fit}   &    \textbf{Best-fit} \\
  &  &  \textbf{(bkg-only)}  & \textbf{(sig+bkg)} \\
Radiogenic n & $ 1.2 ^{+1.0}_{-1.0} $ & $ 1.2 ^{+0.9}_{-0.9} $ & $ 1.2 ^{+0.9}_{-0.9} $ \\ [0.2em]
Surface $\alpha$ & $ 7.4 ^{+15.2}_{-15.4} $ & $ 0.0 ^{+3.1}_{-3.1} $ & $ 0.0 ^{+1.8}_{-1.8} $ \\ [0.2em]
Shadowed $\alpha$ & $ 15.2 ^{+12.3}_{-10.3} $ & $ 18.3 ^{+8.9}_{-7.4} $ & $ 16.5 ^{+8.2}_{-6.9} $ \\ [0.2em]
Dust $\alpha$ & $ 44.2 ^{+39.9}_{-40.4} $ & $ 91.6 ^{+10.6}_{-11.9} $ & $ 64.1 ^{+15.9}_{-16.2} $ \\ [0.2em]
\hline
\textbf{Total Bkg} & $ 68.0 ^{+49.9}_{-45.2} $ & $ 111.1 ^{+16.3}_{-15.2} $ & $ 81.7 ^{+20.3}_{-18.0} $ \\ [0.2em]
\hline
\mbox{$M_{\chi} = 1$ TeV/$c^{2}$}    & -- & -- & $ 32.0 ^{+23.9}_{-22.4} $ \\ 
 [0.2em] 
\hline 
\hline
\end{tabular}
\end{center}
\end{table}

Out of the observed data, 21 events are reconstructed by the neural network (NN) machine-learning algorithm (described in Section: \Nameref{subsec:reco_pos})
as originating from above the LAr fill level, with $z_{\rm NN} > 551$~mm above the origin.
This is consistent with the expected number of shadowed $\alpha$-decay backgrounds, and with the PLR fit results.
For computational reasons it was not possible to include this cut as part of the event selection; instead we mention this observation as a cross-check of the shadowed $\alpha$-decay background normalization.

\begin{figure}[t!b]
 \centering
 \includegraphics[width=1.\linewidth]{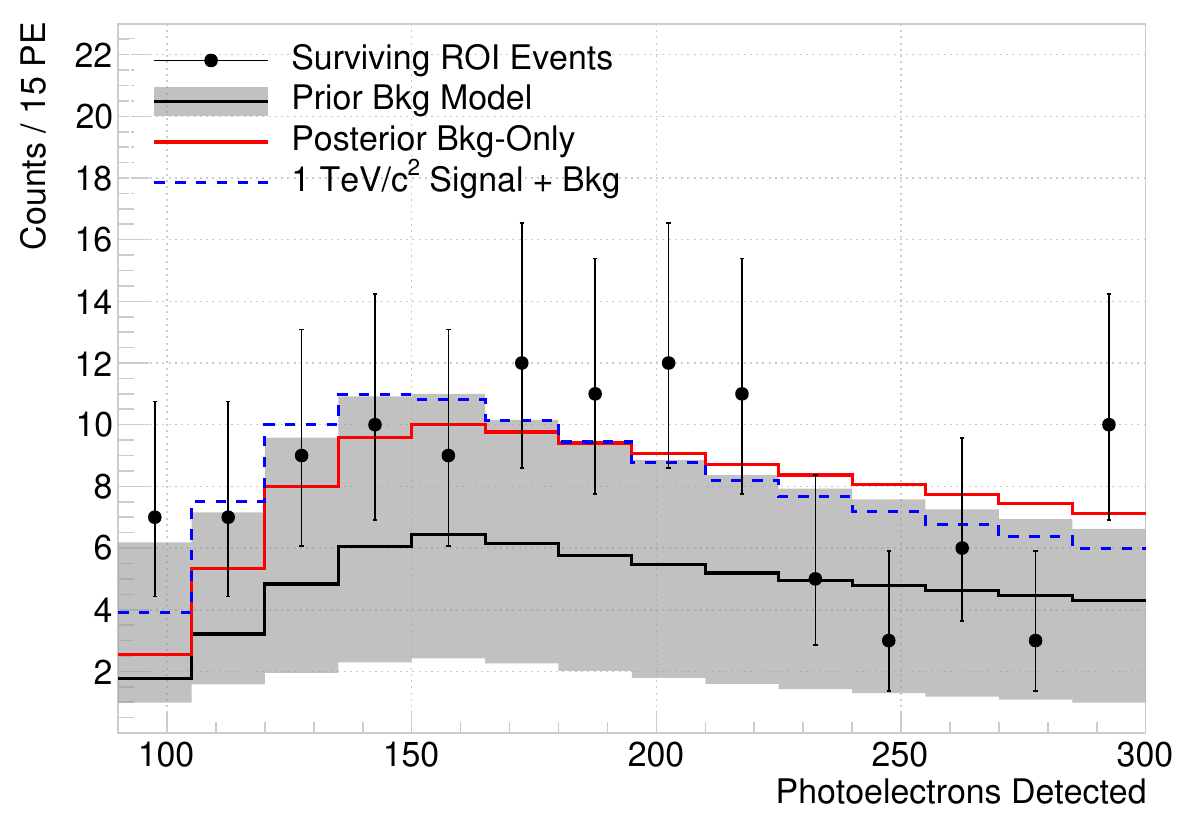}
 \includegraphics[width=1.\linewidth]{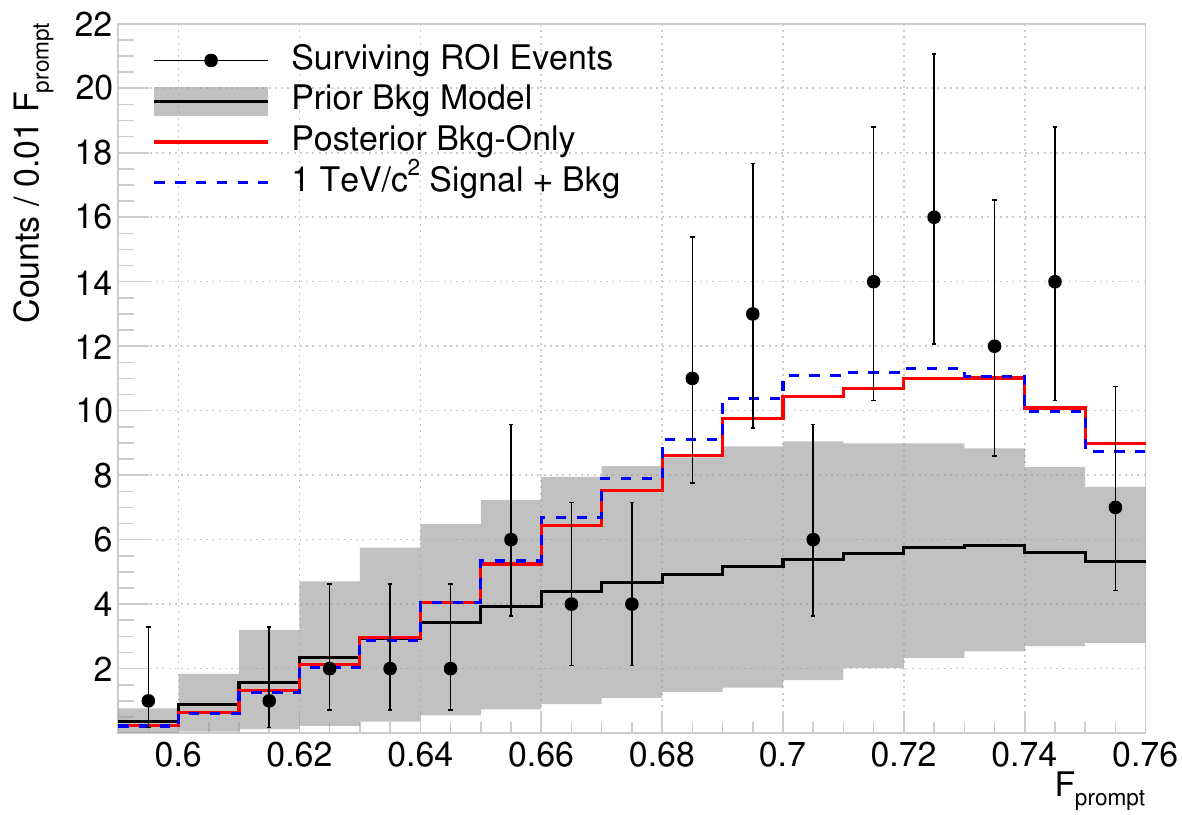}
 \includegraphics[width=1.\linewidth]{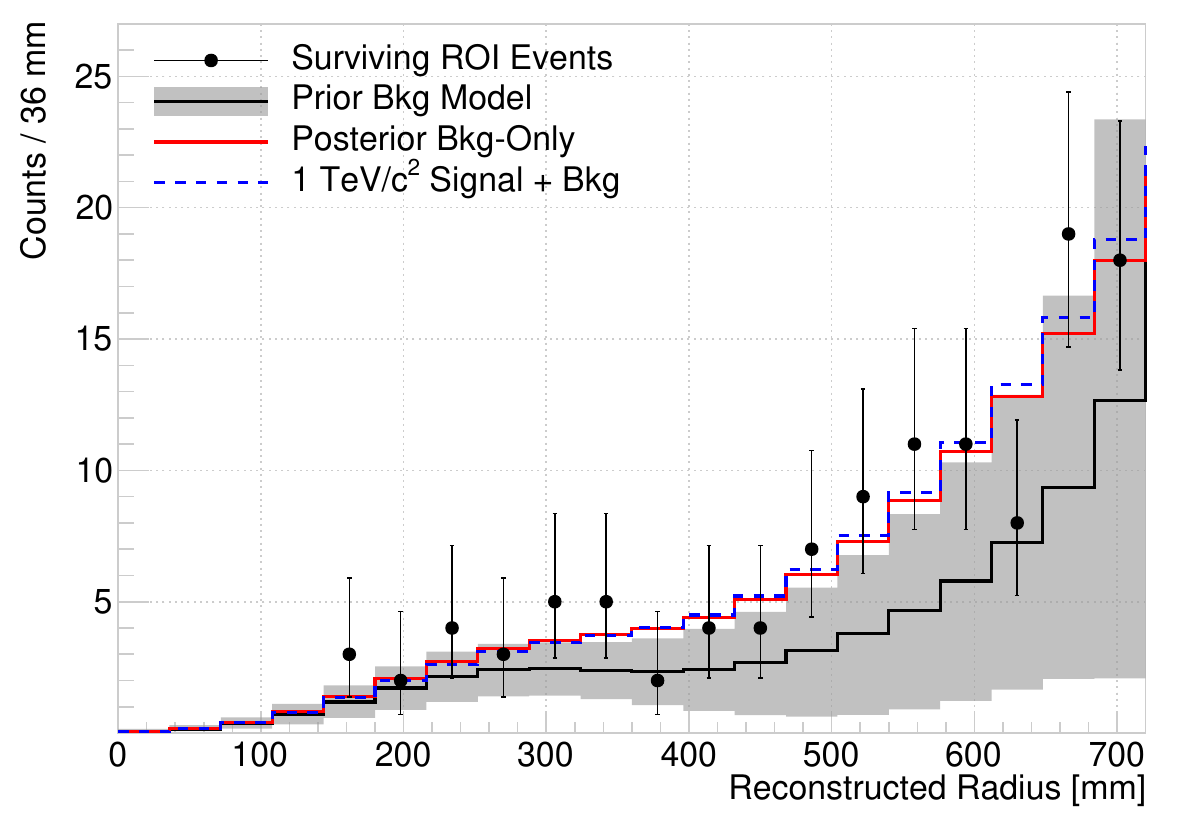}
 \caption{Distributions in \PE\ (top), \FPrompt\ (middle) and \R\ (bottom) of the events observed in the \WIMP\ \ROI, superimposed with the prior background model with $\pm1\sigma$ systematic uncertainty bands (grey), the best-fit background model under the background only hypothesis (red), and the best-fit signal+background model under the \mbox{$M_{\chi} = 1$ TeV/$c^{2}$} hypothesis (blue). Error bars shown on the data points are the 68\% confidence intervals around the bin contents using the Pearson $\chi^{2}$ distribution.
 }
 \label{F:DataVsPriorModel}
\end{figure}

Figure~\ref{F:DataVsPriorModel} shows 1D histograms in the three observables (\PE,\ \FPrompt, and \R) of the data events passing all selection criteria, superimposed with the prior background model with $\pm1\sigma$ systematic uncertainty bands, the best-fit background model under the background-only hypothesis, and the best-fit signal+background model under the \mbox{$M_{\chi} = 1$ TeV/$c^{2}$} hypothesis. 
The excess of events observed above the prior background model can be accommodated by an excess in the dust $\alpha$-decay background, in a manner that is also consistent with the signal+background hypothesis.

In the signal+background hypothesis, the \WIMP\ cross section parameter is allowed to float into the negative range; when it converges to a negative value, the best-fit number of \WIMP\ events is considered to be zero, consistent with the background-only hypothesis. The best-fit WIMP cross section is positive for all $M_{\chi}>40$~GeV/$c^{2}$, while a small negative best-fit cross section is found for $M_{\chi} \le 40$~GeV/$c^{2}$.

\begin{figure}[!htbp]
 \centering
 \includegraphics[width=1.\linewidth]{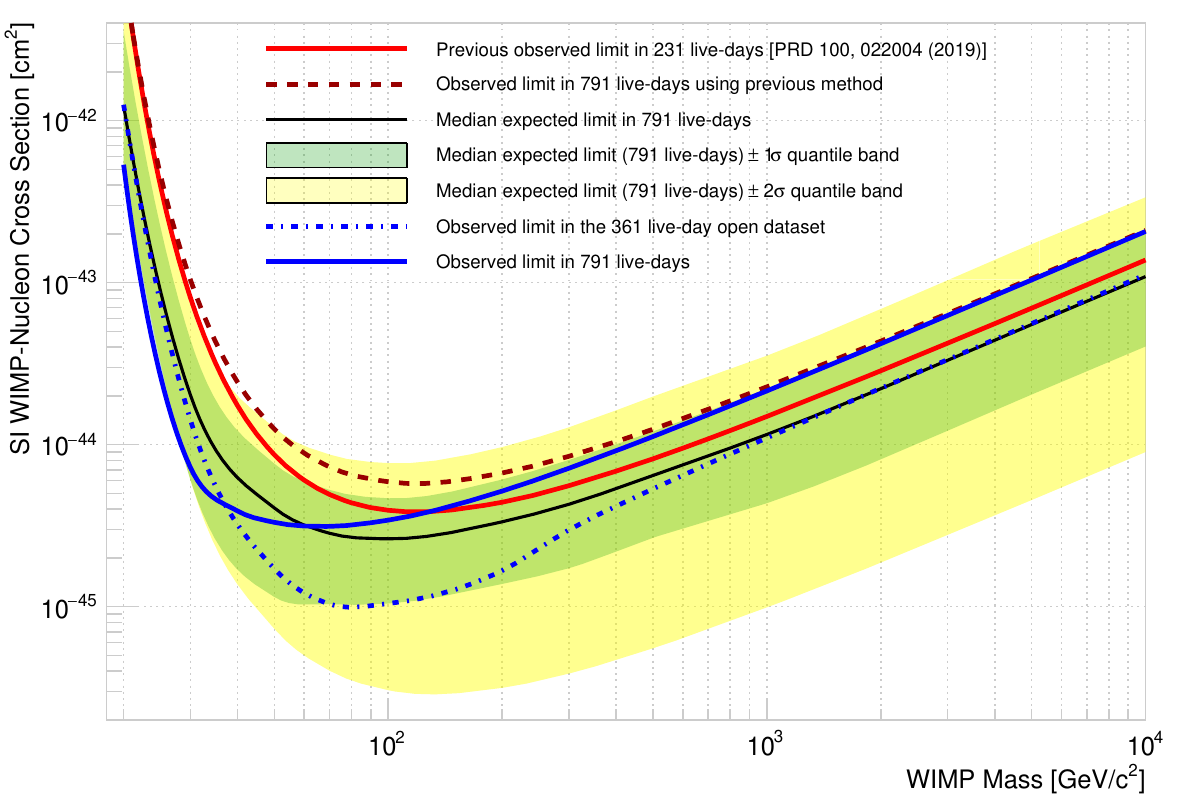}
 \caption{Observed 90\% \CL\ upper limit on the WIMP-nucleon spin-independent cross section as a function of WIMP mass (solid blue) power constrained to the 0.16 quantile (-1$\sigma$). Also shown are the median expected limit (solid black), the $\pm 1\sigma$ and $\pm 2\sigma$ quantile bands (green and yellow respectively), the observed 90\% \CL\ upper limit on the open dataset (dashed blue), the previous observed result from \DEAP\  (solid red) and the result of repeating the previous analysis on the full dataset (dashed red).
 One additional constraint is applied: for each WIMP mass, the minimum value of the upper limit is never allowed go below 2.3~signal events excluded at 90\% \CL.
 }
 \label{F:FinalLimit}
\end{figure}

A one-sided test statistic is used to set a 90\% \CL\ upper limit on the WIMP-nucleon spin-independent cross section as a function of WIMP mass.
The results are shown in Figure~\ref{F:FinalLimit}. The asymptotic approximation~\cite{cowan2011asymptotic} is used to calculate the observed upper limit; this was demonstrated to be valid for our scenario by generating the PLR test statistic distribution for 10,000 pseudo-datasets drawn from the prior background model and verifying that the distribution followed the expected asymptotic approximation.
Comparing the observed value of the log-likelihood in the background-only hypothesis to the distribution from pseudo-datasets yields a $p$-value of 15\%.  
The posterior background-only model is compatible with the observed data.

The expected limit is calculated for each WIMP mass by generating and fitting background-only toy experiments with the \PLR\ analysis. For each toy experiment, the 90\% \CL\ upper limit is computed, and the distribution of upper limits on cross section is plotted in a histogram. From this distribution, the median expected limit is shown on Figure~\ref{F:FinalLimit} as well as the $\pm 1\sigma$ ($\pm 2\sigma$) bands corresponding to the 0.16/0.84 (0.025/0.975) quantiles of the distribution. 

At WIMP mass values in the range 20--60~GeV/$c^{2}$, a better limit is observed compared to the median expected limit. This is a result of a downward fluctuation of the number of observed \ROI\ events with respect to the expected background specifically in the low-PE region of the parameter space that coincides with the signal model for these low WIMP masses. As per the recommendations set out in Ref.~\cite{statswhitepaper}, the observed limit is power-constrained at the 0.16 quantile (-1$\sigma$). Further, the minimum value of the upper limits are constrained to never go below 2.3~signal events excluded at 90\% \CL.

With the full datset, we set an upper limit of \PaperThreeWIMPLimitOneHundredGeV\ at a \WIMP\ mass of 100~GeV/$c^{2}$, and of \PaperThreeWIMPLimitOneTeV\ at 1~TeV/$c^{2}$. 
Due to the excess of data observed in the ROI, these limits are weaker than the median expected limits of \PaperThreeExpectedWIMPLimitOneHundredGeV\ at 100~GeV/$c^{2}$, and of \PaperThreeExpectedWIMPLimitOneTeV\ at 1~TeV/$c^{2}$ for \PaperThreeLiveTimeNum~live-days.
The limit at 1~TeV/$c^{2}$ reported here is also weaker than the previously observed limit of Ref.~\cite{deapdarkmatter2019}.
The minimum of the observed limit curve is \PaperThreeWIMPLimitSixtyGeV\ 
at a WIMP mass of 60~GeV/$c^{2}$.

For \mbox{$M_{\chi} \ge 40$~GeV/$c^{2}$}, significantly better limits are observed when considering the open dataset only, as shown in Figure~\ref{F:FinalLimit}.  The excess of data observed over the background prediction in the blind part of the dataset explains this discrepancy.

Finally, we observe that repeating the cut-and-count analysis of Ref.~\cite{deapdarkmatter2019} on the full dataset, applying the stricter event selection including a narrower ROI that was used in this previous version of the DEAP-3600 dark matter search, yields 11 events: 2~in the open dataset and 9~in the blind dataset.
The background estimate from our updated model, scaled up in livetime and including the expectation from dust $\alpha$-decay events, yields a prediction of 5.0 events, for this event selection, so the $p$-value for observing 11 events or more is 1.3\%.
The corresponding limit curve is also shown on Figure~\ref{F:FinalLimit}.
Having obtained a better limit in Ref.~\cite{deapdarkmatter2019} is due to the observation of no data point in the ROI at the time;
given our updated model expecting 1.5 background events in \PaperTwoLiveTimeNum~live-days, the $p$-value for having observed zero event then is 22\%.

Hardware upgrades to the DEAP-3600 detector have been installed to mitigate the sources of background that limit the sensitivity to WIMP dark matter in the 2016--2020 dataset shown here.  The upgrades feature: pyrene-doped polystyrene-coated flowguides~\cite{DEAP_Pyrene,DEAP_Pyrene_Fluorescence,garg_2023} to shift $\alpha$-decay background events out of the WIMP ROI, an external cooling system to prevent LAr condensation on acrylic in the neck, and a dust-filtration pipe to remove particulates and recirculate LAr.

\section*{Methods}\setcurrentname{Methods}\label{sec:method}

\subsection*{DEAP-3600 Detector Overview}\setcurrentname{DEAP-3600 Detector Overview}\label{sec:detector}

\DEAP\ is a single-phase \LAr\ dark matter detector, located approximately 2 km underground in the \SNOLAB\ laboratory in Sudbury, Ontario, Canada. The detector design is summarised here; a full description of the detector design is in Ref.~\cite{amaudruz_design_2017}.

Ultrapure LAr of target mass \larmasserror\ \cite{deap_39Ar_specific_activity} is contained inside a spherical, radiopure acrylic vessel (\AV) with an inner diameter of \AVInnerDiameter\ and a thickness of \AVThickness. Gaseous argon (\GAr\/) occupies the remaining inner volume above the \LAr\ fill level, located \larlevelcm\ above the equator. On the inner surface of the acrylic vessel resides a \TPBThickness\/-thick layer of tetraphenyl butadiene (\TPB\/) wavelength shifter, which shifts ultraviolet (\UV) \ArWaveLength\ scintillation light from \LAr\ into the visible region of the electromagnetic spectrum ($\sim$ \TPBWaveLength\/). Visible photons are detected by 255 inward-facing Hamamatsu photomultiplier tubes (\PMTs\/)~\cite{the_deap_collaboration_-situ_2017}, each optically coupled to the end of a \LGLength\/-long cylindrical acrylic light guide bonded to the acrylic vessel. Interspersed between the light guides are 486 high-density polyethylene and polystyrene filler blocks, which provide thermal insulation to the \PMTs\ which are operated at approximately room temperature. The filler blocks also shield the inner volume from radiogenic neutron backgrounds originating from the \PMTs\/. 

The inner volume is accessed by an opening at the top of the acrylic vessel that connects to an acrylic neck. Inside the neck resides a stainless steel cooling coil filled with liquid nitrogen (\LN\/) and two acrylic flowguides, the ``inner'' flowguide and ``outer'' flowguide. \LAr\ produced from condensed \GAr\ inside the neck is guided into the inner volume by the two flowguides, which also regulate the flow of warmer \GAr\ back up into the neck to be cooled. 
Four small ``neck veto \PMTs'' are coupled to scintillating fibers on the outside of the acrylic neck of the detector, to detect light originating from background events occurring there.

The inner detector components are all housed inside a spherical steel shell that is submerged inside a large, cylindrical tank of \MuonVetoDiameter\ diameter and \MuonVetoHeight\ height filled with ultrapure water. Attached to the outside of the steel shell are 48 outward-facing \PMTs\/, which in conjunction with the water tank comprise the Cherenkov muon veto used to tag and reject cosmogenic muon backgrounds. Ultrapure water also acts as a buffer to suppress neutron backgrounds from the rock surrounding the experiment.

\subsection*{Detector Electronics and Data Acquisition}\label{sec:detector_daq}

All of the detector electronics sit on the deck located above the water tank, housed in 3 computer racks. The electronics are comprised of 3 separate systems: the front end system, the Digitizer and Trigger Module (\DTM) and the Data Acquisition System (\DAQ), based on the \MIDAS\ software infrastructure for event readout \cite{lindner_deap-3600_2015}. Together the electronics are responsible for the digitization, triggering and acquisition of signals from the \PMTs.

The front end system consists of a high voltage supply for the \PMTs\ and 27 signal conditioning boards (\SCBs)
responsible for the 255 \LAr\ \PMTs, the 48 muon veto \PMTs\ and the 4 neck veto \PMTs, each \SCB\ being connected to up to 12 \PMTs. 
The \SCBs\ shape the \PMT\ signals and send them to high-gain channels (HGCs: 32~CAEN V1720s with \FastDigitizerRate\ sampling rate)
and low-gain channels (LGCs: 4~CAEN V1740s with \SlowDigitizerRate\ sampling rate) for amplification.
The HGCs are the default readout channel, while the data saved by the LGC provide a way to correct digitizer clipping effects in the HGC for large pulses.

The HGC from the LAr PMTs are summed together and the resulting waveform is digitized and analyzed by the DTM to make a trigger decision.
The trigger algorithm used by the \DTM\ is described in \cite{deapdarkmatter2019}. 
If the event passes the trigger condition set by the \DTM, the digitized \PMT\ signals are read out for \DAQWindow\ by the \DAQ\ system comprised of 6 front end computers and 1 main computer. The main computer runs the event builder and the logger program, which ``builds'' and compresses the event before writing it to disk. 
Events are saved in binary files using the \MIDAS\ format, 
and are then processed with the Reactor Analysis Tool (\RAT)~\cite{rat_github} to obtain \PMT\ waveforms upon which event reconstruction algorithms are run. 
These reconstructed event-by-event observables are recorded in \CERNRoot\ files~\cite{brun_root_1997} for offline data analyses. 

The RAT framework is also used to produce MC simulations of events in \DEAP\ by interfacing with GEANT4~\cite{Agostinelli:2003fg}.
Our simulation of the \DAQ\ is then run on these events, followed by the same event reconstruction algorithms as are used on real data.

\subsection*{Event Reconstruction}\setcurrentname{Event Reconstruction}\label{sec:reconstruction}
This analysis considers three main observables for each event: the number of detected \PE\ as an event energy estimator, the \PSD\ parameter \FPrompt\ and the reconstructed event vertex radial position \R. 
One additional variable used in the analysis is \PIFGAr\ the number of \PMTs\ looking into the \LAr\ volume that detected a pulse before any pulse was detected in a \PMT\ looking above the \LAr\ fill level. This variable was specifically developed to use for background rejection against shadowed $\alpha$-decays originating from the neck of the detector, for which \PIFGAr\ is low. In this section, these four analysis variables are described in more detail.

\subsubsection*{Energy Estimation}\setcurrentname{Energy Estimation}\label{subsec:reco_charge}

In DEAP-3600, the energy of an event is characterised by the amount of charge it induces in the \PMTs\ in a \DAQWindowRec\/-long event window following the trigger time.
An algorithm is used to correct for saturation effects such as digitizer clipping and non-linear \PMT\ response.
Individual pulses with $> 150$~PE typically result in clipping in the HGC, while the LGC remain unclipped at this scale.
For clipped pulses, the charge recorded by the HGC is corrected based on a function that was obtained by comparing HGC pulses to LGC pulses over a large data sample.
This pulse charge estimate is subsequently adjusted for \PMT\ nonlinearity using a model which numerically emulates dynode space charge effects.

From this input, the number of detected \PE\ in each pulse is determined using a Bayesian counting algorithm~\cite{deapdarkmatter2019, burghardt_2018, butcher_2015}, designed to remove \PMT\ noise contributions such as afterpulsing, which would otherwise degrade the energy resolution.
For each \PMT\ pulse recorded in the first \DAQWindowRec\ of the event waveform, the algorithm calculates the likelihood distribution for the number of \PE\ produced specifically from \LAr\ scintillation. Probabilities are assigned for each detected pulse to originate from scintillation as well as other \PMT\ effects, such as dark rate, late and double pulsing, early pulsing and afterpulsing using the corresponding input prior distributions based on the LAr scintillation time profile and the afterpulsing charge-time distribution for the \PMT\ in question. The posterior distribution is calculated using a combination of the prior distributions and the single \PE\ charge distribution.  The mean of this posterior distribution is taken as the estimated number of \PE\ generated from \LAr\ scintillation. This quantity is summed for all pulses across all 255 \LAr\ \PMTs\/ to estimate the total number of \PE\ produced in the event.

\begin{figure}[htb]
 \centering
 \includegraphics[width=1.01\linewidth]{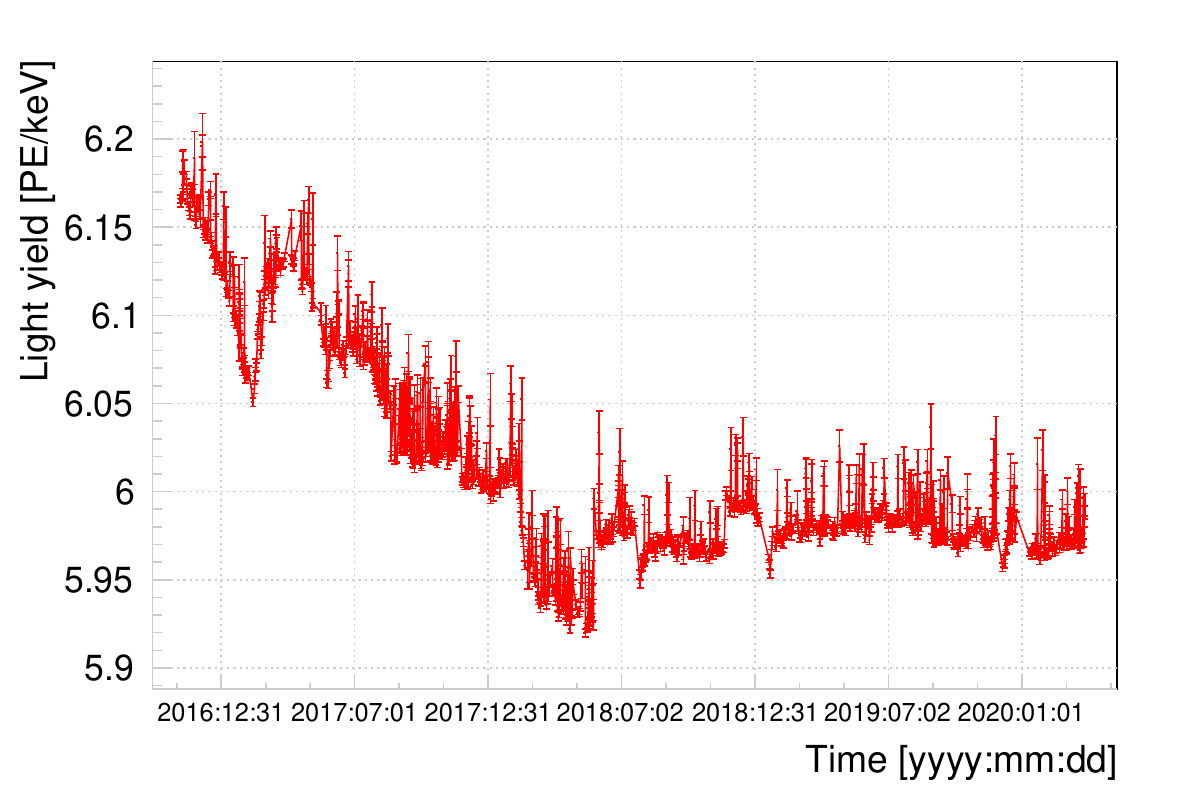}
 \caption{Light yield in terms of PE from the Bayesian counting algorithm designed to remove PMT noise contributions such as afterpulsing, as a function of date across the dataset, determined from calibration fits to the $^{39}$Ar spectrum.}
 \label{F:LY_vs_RunNumber}
\end{figure}

The light yield of the detector slowly decreased in the first year of the dataset, before it stabilised~\cite{deap_39Ar_halflife}.
To account for this shift, the data used in this analysis are time-calibrated to a specific run where the value of the light yield is measured to be \SI{6.1}{PE/keV}. 
Using an interpolated curve of the light yield as a function of time, shown in Figure~\ref{F:LY_vs_RunNumber}, the \PE\ values are shifted to account for the difference in the light yield at the time the event was observed with respect to the reference light yield.

\subsubsection*{Pulse-Shape Discrimination}\setcurrentname{Pulse-Shape Discrimination}\label{subsec:reco_psd}

The \PSD\ parameter, called \FPrompt, is defined as the ratio of the number of prompt \PE\ detected by the \PMTs\ within the prompt window [\SI{-28}{ns}, \SI{60}{ns}] of the event time to the total number of \PE\ counted within [\SI{-28}{ns}, \DAQWindowRec] of the event time:

\begin{equation}
\FPrompt = \frac{\sum_{t=\SI{-28}{\ns}}^{\SI{60}{\ns}} \PE(t)}{\sum_{t=\SI{-28}{\ns}}^{\SI{10}{\us}} \PE(t)}.
\label{eq:fprompteqn}
\end{equation}

In Ref.~\cite{DEAP_PSD} this choice of prompt window was shown to maximally discriminate \NRs\ from \ERs\ in \DEAP.

\subsubsection*{Position Reconstruction}\setcurrentname{Position Reconstruction}\label{subsec:reco_pos}

The main position reconstruction algorithm used in this analysis is the same \PE-based algorithm as in Ref.~\cite{deapdarkmatter2019}.
We consider a position likelihood function $\mathcal{L}(\vec{x})$ that describes the probability of observing an event at a given vertex $\vec{x} = (x,y,z)$:

\begin{equation}
	\mathcal{L}(\vec{x}) = \prod_{i=1}^{N_{\mathrm{PMT}}}\mathrm{Pois}(n_{i}|\lambda_{i}).
\end{equation}
The $z$ axis is the vertical axis collinear with the \AV\ neck, while $x$ and $y$ axes are on the equatorial plane, with the centre of the detector being located at the origin of the coordinate system. 
For a PMT $i$ located at position $\vec{r}_{i} = (x_{i},y_{i},z_{i})$, $\mathrm{Pois}(n_{i}|\lambda_{i})$ is the Poisson probability of detecting $n_{i}$ \PE\ within the \DAQWindowRec\ event window.
The expected number of \PE\ detected by PMT $i$, $\lambda_{i}$, is dependent on the magnitude $|\vec{x}|$, the angle between $\vec{x}$ and $\vec{r}_{i}$ and the total number of observed \PE\ across all \PMTs. 
The likelihood function refers to look-up tables to find the value of $\lambda_{i}$ for a given $\vec{x}$. 
These look-up tables are generated using high-statistics MC simulations of \LAr\ scintillation events produced in the detector at discrete positions along the $(x,y,z)$ axes.
The reconstructed position returned by the algorithm is given by the value of $\vec{x}$ that maximises $\mathcal{L}(\vec{x})$. 
The variable used in the \PLR\ analysis is the reconstructed radius \R\/, determined using the reconstructed Cartesian coordinates as 
$\R = \sqrt{x_{\rm rec}^{2} + y_{\rm rec}^{2} + z_{\rm rec}^{2}}$.
The performance of this algorithm improved after re-calculating the look-up tables following updates to our \LAr\ optical model discussed in Section: \Nameref{sec:systematics}.

As described in Refs.~\cite{deapdarkmatter2019,deapPosRec}, \DEAP\ also employs a time-residual based position reconstruction algorithm that uses both the \PE\ and time information of the early pulses detected to reconstruct the event position and event time.
First, for a grid of test positions $\vec{x}_j$ in the \LAr\ target, for each \PMT\ position $\vec{r}_i$ the time-residual distribution $\mathcal{L}^{t\ {\rm res.}}(\Delta t; \vec{x}_j, \vec{r}_i)$ is constructed, where $\Delta t$ is the photon time of flight in \LAr\ from $\vec{x}_j$ to $\vec{r}_i$ based on a simplified optical model taking into account the group velocities of UV photons (emitted by \LAr) and visible photons (emitted by \TPB), plus time delays sampled from exponential distributions with the \LAr\ scintillation and \TPB\ fluorescence time constants, as well as the uncertainties on photon detection time due to light propagation in the light guides and the PMT response time. Then, at the time of event reconstruction, for each \PE\ detected at time $t_i$ in \PMT~$i$ within the first \SI{40}{ns} of the event, the likelihood of a given event position $\vec{x}$ and event time $t_0$ is given by:

\begin{equation}
	\mathcal{L}(\vec{x}, t_0) = \prod_{i=1}^{N_{\mathrm{PE}}}\mathcal{L}^{t\ {\rm res.}}(t_i - t_0; \vec{x}, \vec{r}_i).
\end{equation}
This time-residual based algorithm returns the event time and position that maximize $\mathcal{L}(\vec{x}, t_0)$.
It is used in conjunction with the \PE-based algorithm to evaluate position reconstruction consistency observables that are based on the $\Delta z$ distance and the 3D distance between the two reconstructed vertex positions.  Position reconstruction consistency enters the \WIMP\ search event selection primarily to reduce the shadowed $\alpha$-decay backgrounds from the neck.

A third position reconstruction algorithm employs machine learning, by means of three feed-forward neural networks (FFNN) that each map the observed number of PE detected by the 255 PMTs to one of the three reconstructed position components $x_{\rm NN}, y_{\rm NN}$ and $z_{\rm NN}$.
The network architectures, training methodology and performance validation are detailed in Ref.~\cite{deapPosRec}.
By virtue of being trained on simulated shadowed $\alpha$-decay backgrounds from the neck of \DEAP\ in addition to simulated $^{40}$Ar NR and $^{39}$Ar ER events, this algorithm has good accuracy in reconstructing the position of these three event types, the $z_{\rm NN}$ coordinate thereby providing an additional handle against shadowed events from the neck.
Because this background discrimination power comes at the cost of a slightly worse position resolution for events that truly occur in the LAr volume, the PE-based maximum-likelihood algorithm remains our primary position reconstruction method.

\subsubsection*{Additional Variable Against Shadowed Events from the Neck}\setcurrentname{Additional Variable Against Shadowed Events from the Neck}\label{subsec:reco_pifgar}

The variable \PIFGAr\ is a useful discriminant against shadowed $\alpha$-decay events from the neck of \DEAP.
Because UV photons travel faster in \GAr\ than in \LAr, scintillation photons originating above the fill level can reflect off the \LAr\ surface and then generate early pulses in \PMTs\ looking into the \GAr\ volume, i.e. \PIFGAr\ is lower for events originating in the neck.
In contrast, in events where scintillation is genuinely produced within the \LAr\ volume, a higher number of \PMTs\ looking into the \LAr\ volume detect pulses before the first pulse is detected in a \PMT\ looking above the fill level.

\begin{figure}[htb]
 \centering
 \includegraphics[width=1.01\linewidth]{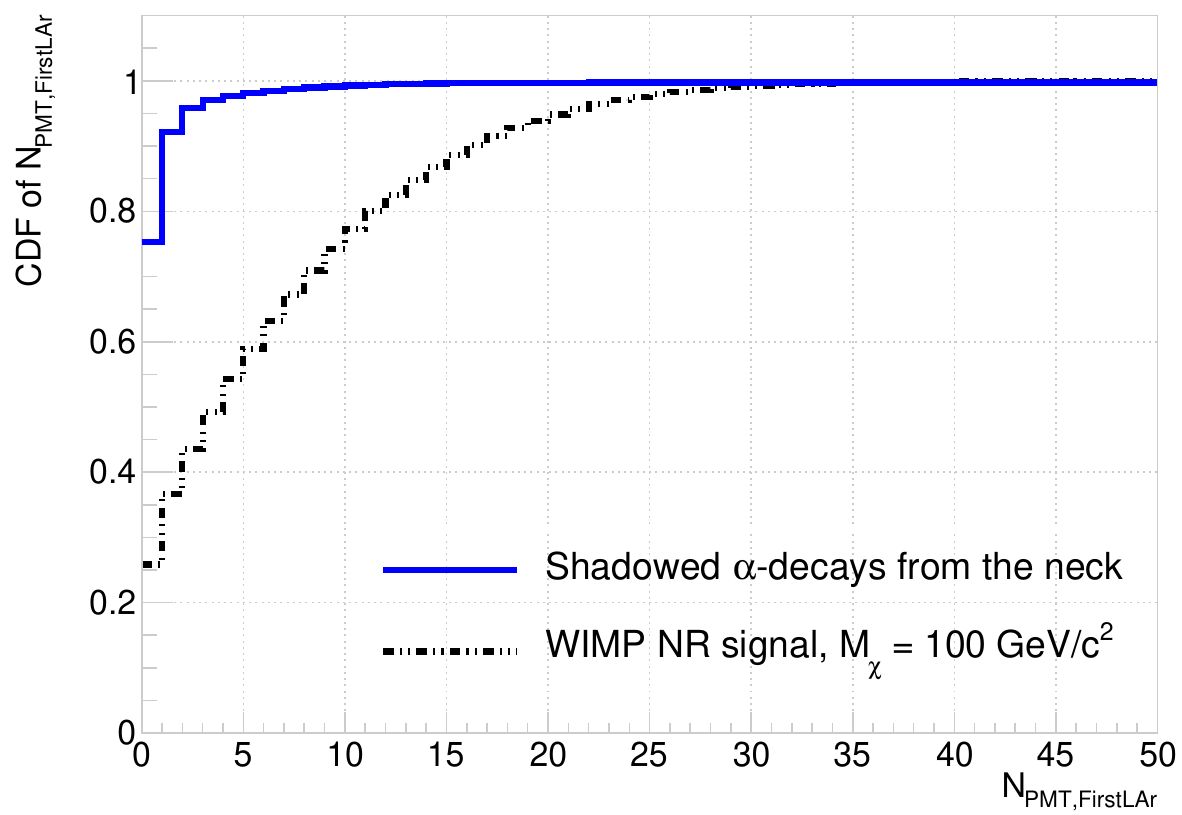}
 \caption{Cumulative distribution functions (CDFs) of \PIFGAr\ from simulated shadowed $\alpha$-decays from the neck and simulated WIMP-induced $^{40}$Ar \NRs.}
 \label{F:PIFGArNeckWIMP}
\end{figure}

Figure~\ref{F:PIFGArNeckWIMP} illustrates the \PIFGAr\ distribution from simulated shadowed $\alpha$-decays from the inner and outer surfaces of the inner flowguide and the inner surface of the outer flowguide in the neck of the detector, and simulated WIMP-induced $^{40}$Ar NR events. The outer surface of the outer flowguide is coupled to the wall of the neck and thus has no direct line of sight to the inner detector volume to produce shadowed $\alpha$-decay events. 

In the analysis of Ref.~\cite{deapdarkmatter2019}, rejecting events with \mbox{\PIFGAr\ $\leq 2$} resulted in a predicted background leakage from shadowed $\alpha$-decays of \mbox{$< 0.5$ events} expected in \PaperTwoLiveTimeNum\ live-days inside the \WIMP\ \ROI.
While this cut is very efficient at removing shadowed $\alpha$-decay events, this results in an important loss of signal acceptance inside the \WIMP\ \ROI.
To leverage the discrimination power of this variable while mitigating the acceptance loss that would arise from cutting on it, instead the \PLR\ fit is run in two ``bins'' defined by \mbox{\PIFGAr $\leq 2$} and \mbox{\PIFGAr\ $> 2$}.
These two bins are chosen based on a compromise between optimal discrimination power between signal and background events from shadowed $\alpha$-decay backgrounds and the computing time required to construct the models to be used in the \PLR\ analysis.

\subsection*{The Profile Likelihood Ratio Method}\setcurrentname{The Profile Likelihood Ratio Method}\label{sec:Lfunc}

The \PLR\ approach is a statistical technique favoured by experiments with a non-negligible background rate as a method of improving their signal sensitivity, using background and signal distributions as a method of background subtraction.
In this analysis, the likelihood function is comprised of two separate terms:

\begin{equation}
\begin{split}
    \mathcal{L}(\sigma_\chi | \NuisancePars\/) =\ & \mathcal{L}_{\mathrm{Model}}(\sigma_\chi | \NuisancePars\/)\ \times \mathcal{L}_{\mathrm{Constraint}}(\NuisancePars\/),
    \label{E:LikelihoodDEAP}
    \end{split}
\end{equation}
where $\sigma_\chi$ is the WIMP-nucleon spin-independent cross section,
\mbox{$\NuisancePars\ = \{\theta_{1}, \theta_{2},.., \theta_{N}\}$} are a set of $N$ nuisance parameters, 
$\mathcal{L}_{\mathrm{Model}}(\sigma_\chi | \NuisancePars\/)$ is the main model term and
$\mathcal{L}_{\mathrm{Constraint}}(\NuisancePars\/)$ is the constraint term. 

The PLR is then defined as the ratio of the conditional maximum likelihood (for a given $\sigma_\chi$) to the unconditional maximum likelihood (where $\sigma_\chi$ is allowed to float).

\subsubsection*{Model Term}\setcurrentname{Model Term}\label{subsec:Lfunc_PDF}

The main term in the likelihood function, $\mathcal{L}_{\mathrm{Model}}(\sigma_\chi | \NuisancePars\/)$, is an un-binned term calculated from multi-dimensional probability distribution functions (PDFs) of observables in the WIMP signal model and each background model. The choice of observables taken as model parameters are motivated foremost by their discrimination ability between signal and background events. The full term is given by:

\begin{equation}
\begin{split}
    \mathcal{L}_{\mathrm{Model}}(\sigma_\chi;\{\theta \}) =\ & \mathrm{Pois}(N_{\mathrm{obs}}|N_{{\mathrm{exp}}})\ \times \\ & \prod_{j=1}^{N_{\mathrm{obs}}}\Big(\sum_{i=1}^{N_{\mathrm{Models}}}\frac{N_{\mathrm{exp,i}}}{N_{\mathrm{exp}}} \times f_{i|j}\{\theta \}) \Big),
    \label{E:PDFTerm}
\end{split}
\end{equation}
where $f_{i|j}\{\theta \}$ is the PDF evaluation for model $i$, for observed event $j$ given its values of \PE, \FPrompt, \R, and \PIFGAr; 
$N_{\mathrm{exp}}$ is the total number of expected signal and background events inside the ROI in the dataset and 
$N_{\mathrm{obs}}$ is the total number of observed events inside the ROI in the dataset. 

In other words, the sum computed inside of the circular brackets in Equation \ref{E:PDFTerm} represents the total likelihood for observed event $j$ to reside inside the ROI, given its \PE\/, \FPrompt\/, \R\ and \PIFGAr\ values. The total term is then given by the product over all observed events $N_{\mathrm{obs}}$, multiplied by a Poisson term comparing the number of expected events with the number of observed events.

\subsubsection*{Constraint Term}\setcurrentname{Constraint Term}\label{subsec:Lfunc_con}

Systematic uncertainties are incorporated into the analysis in the form of nuisance parameters, which are constrained in the likelihood function by the constraint term, $\mathcal{L}_{\mathrm{Constraint}}(\NuisancePars)$. The total constraint term is given by the product of the individual constraint terms for each nuisance parameter:

\begin{equation}
    \mathcal{L}_{\mathrm{Constraint}}(\{\theta \}) = \prod_{k=1}^{N_{\theta}}f(\theta_{k}),
    \label{E:ConTerm}
\end{equation}
where $N_{\theta}$ is the total number of nuisance parameters. 

Each nuisance parameter has a corresponding constraint PDF, denoted $f(\theta_{k})$, which contributes to the
likelihood function with the appropriate level of systematic uncertainty. Parameters with Gaussian uncertainties have
Gaussian constraint PDFs of the form $\mathrm{Gaus}(\mu_k,\sigma_k)$, where $\mu_k$ corresponds to the nominal value of
the nuisance parameter and $\sigma_k$ is the standard error corresponding to the uncertainty. The nominal value of each
nuisance parameter and the corresponding uncertainties used by each constraint PDF are derived from ex-situ
measurements made either from high-energy sidebands located outside of the ROI or from separate calibration datasets;
these are detailed in Section~\Nameref{sec:systematics}.

\subsection*{Signal and Background Model Implementation}\setcurrentname{Signal and Background Model Implementation}\label{sec:models}

\subsubsection*{Overview}\setcurrentname{Overview}\label{subsec:models_overview}

The \PLR\ analysis software is written in C++ and uses the MIGRAD algorithm in the Minuit2 library \cite{minuit} of \CERNRoot\ to perform the minimisation of the custom-written likelihood function, which returns the negative log-likelihood. 
As outlined in the previous section, the WIMP signal model and each background model considered in the analysis are described by two sets of multi-dimensional PDFs based on three main event observables, \PE, \FPrompt\ and \R, with each of the two sets modelled separately for different \PIFGAr\ values ($\leq$ 2 and $>$ 2). 

The shapes of the multi-dimensional PDFs are determined using a combination of theoretical models, data observed in CRs and sidebands, simulations, and calibration data. 
One-dimensional PDFs describing the \PE\ distributions of each model are obtained by fitting one-dimensional histograms of the \PE\ distributions with empirical functions. 
Then, two-dimensional PDFs describing the \FPrompt\ and \R\ distributions are obtained by fitting one-dimensional histograms of \FPrompt\ and \R\ with empirical functions in multiple \PE\ bins;
the parameters of these empirical functions are then modelled as a function of \PE, to account for the variations in \FPrompt\ and \R\ as a function of \PE.

In total, there is one signal model for each WIMP mass considered, and four background models considered in this analysis:

\begin{itemize}
\item Argon NRs due to interactions with WIMP dark matter (signal), using as a benchmark the standard halo model~\cite{mccabe_astrophysical_2010} with a local density of \SI{0.3}{GeV/cm^3} and a Maxwell-Boltzmann velocity distribution below an escape speed of \SI{544}{km/s} with $v_0 = 220$~km/s;
\item Radiogenic neutrons coming from the surrounding detector materials, such as the \PMT\ glass, produced primarily through spontaneous fission and ($\alpha$,n) reactions;
\item ``Surface'' $\alpha$-decays from $^{210}$Po on the inner surface of the \AV;
\item ``Shadowed'' $\alpha$-decays from $^{210}$Po on the surfaces of the acrylic flowguides in the neck of the detector;
\item ``Dust'' $\alpha$-decays originating from trace amounts of small dust particulates suspended in the LAr volume.
\end{itemize}

Additionally, there are three event types determined to be negligible in the \WIMP\ ROI:

\begin{itemize}
\item $^{39}$Ar $\beta$-decays originating from within the LAr are negligible by construction of the ROI, based on our PSD model detailed in Ref.~\cite{DEAP_PSD};
\item Cherenkov light produced by $\gamma$-rays in acrylic are determined, by repeating the study from Ref.~\cite{deapdarkmatter2019}, to contribute $< 0.37$ events (90\%~CL) in the ROI used here;
\item Cosmogenic neutron backgrounds produced by high-energy atmospheric muon interactions with the detector and its environment are determined to be negligible in the ROI based on the analysis of Ref.~\cite{deapdarkmatter2019}.
\end{itemize}

The dominant sources of background events in this analysis are dust $\alpha$-decays, described in the following section, and shadowed $\alpha$-decays.
Regarding the latter, as detailed in Ref.~\cite{deapdarkmatter2019}, background events consistent with LAr scintillation from $^{210}$Po $\alpha$-decays on the surfaces of the neck flowguides are problematic for our position reconstruction algorithms due to their complex photon propagation topology involving the GAr-LAr interface and shadowing effects, whereby a large fraction of scintillation photons generated by the decay are absorbed by the acrylic flowguides and thus do not produce any detectable \PE\ at the \PMTs. As a result, shadowed $\alpha$-decays can reconstruct inside the fiducial volume.
A number of the event selection cuts presented in Section~\Nameref{subsec:resultsROI} are used specifically against this background.

As input to this PLR analysis the methods used to constrain background rates using CRs are as described in Ref.~\cite{deapdarkmatter2019}, with the addition of the dust $\alpha$-decay rate estimate using a dedicated CR, and an updated data-driven model for surface $\alpha$-decays.

\subsubsection*{Dust $\alpha$-Decay Background}\setcurrentname{Dust $\alpha$-Decay Background}\label{subsec:models_dust}

\begin{figure}[htb]
 \centering
 \includegraphics[width=1.01\linewidth]{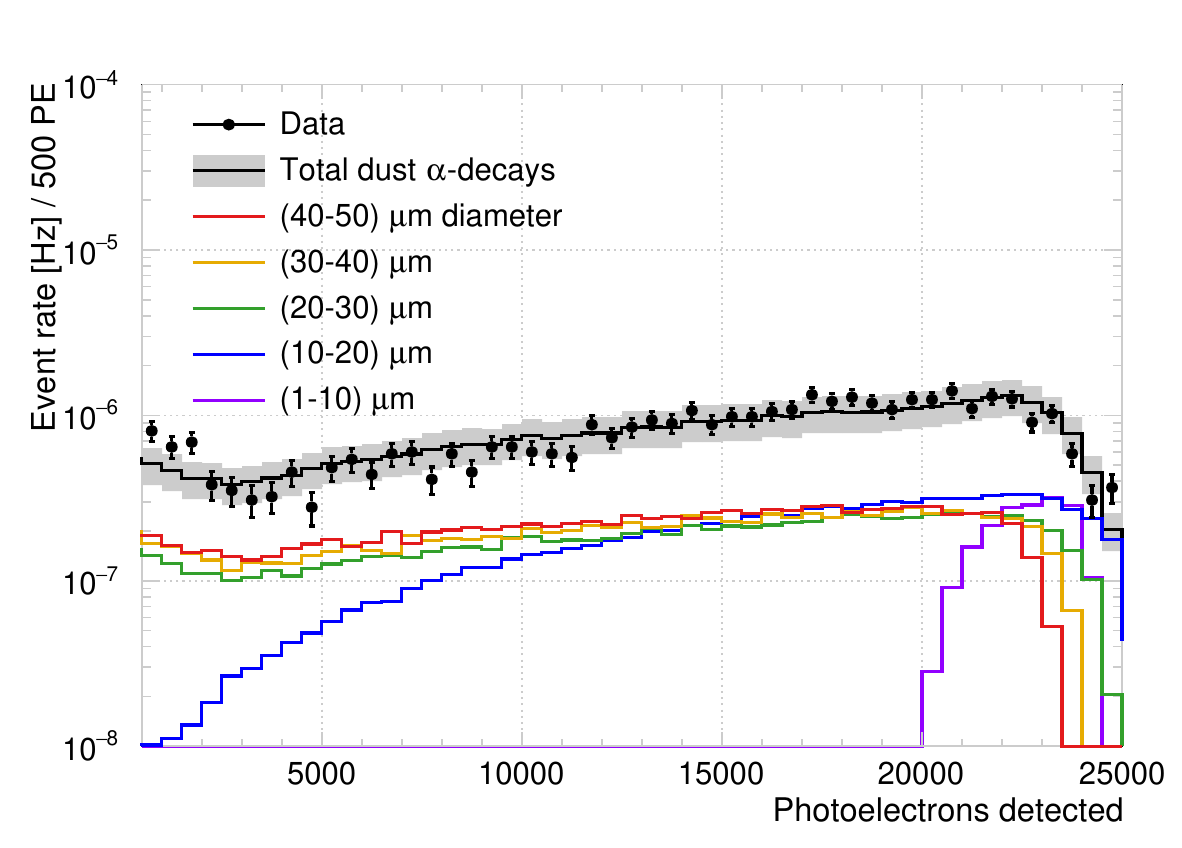}
  \caption{Simulated samples of copper dust $\alpha$-decays fitted to the data in the dust \CR\ using a power function of the dust size.
  }
 \label{ref:Fit_spectrum}
 \end{figure}

An additional background model component is introduced in this analysis: $\alpha$-decays coming from a low level of dust particulate contamination inside the \LAr.
The energy of the $\alpha$ particle from such a decay is attenuated in the dust particulate before reaching the \LAr, and fewer scintillation photons are produced: consequently fewer \PE\ are observed in these events compared to $\alpha$-decay events where the full energy is deposited in \LAr.
As shown in Figure~\ref{ref:Fit_spectrum}, a sample of these events, unexplained by other components of our background model, is seen in the \NR\ band over the range \SI{500}{PE}--\SI{20000}{PE} for \R~$< 720$~mm, hereafter called the ``dust \CR''.
The additional event selection cuts applied in this \CR\ are detailed in Table~\ref{T:Cuts}.
In particular, the cut \mbox{$\PIFGAr > 2$} is kept here in order to remove shadowed $\alpha$-decays from the dust \CR.

This background is consistent with small dust particulates ranging in size from 1~$\mu$m--50~$\mu$m that could have been left in the detector after construction.
A computational fluid dynamics calculation showed that the convective flow in the \LAr\ would effectively mix these particulates uniformly.
This is also consistent with a characterization of the \LAr\ flow patterns inferred from the displacement of coincident $^{214}$Bi and $^{214}$Po decays observed in \DEAP\ data.

Since we are uncertain whether the particulates were introduced before or after applying the TPB coating, we allow coated and uncoated particles in the analysis. 
The particulates are modeled as copper spheres with and without TPB-coating (with 3\,$\mu$m thickness);
they are simulated uniformly throughout the \LAr, with $^{210}$Po $\alpha$-decays generated uniformly within the particulates.
In order to obtain the probability distribution functions of dust $\alpha$-decays, the output of these simulations are fit to observed \PE\ spectrum in the dust \CR, so that the analysis is insensitive to the assumed material making up the dust.
This fit to data constrains the dust $\alpha$-decay rates under the assumption that the distribution of particulate sizes follows a power distribution~\cite{hallman1991establishing, parasuraman2012prediction, akeribLUXZEPLINLZRadioactivity2020}, using the following function:

\begin{equation}
    N_{\rm dust} = N_0 \sum_{i=1}^{5} (D_{\mathrm{upper}}^{p} - D_{\mathrm{lower}}^{p}) h_i
\end{equation}
where $N_0$ is the overall normalization parameter, $p$ is the exponent parameter of the power law, and $D_{\mathrm{lower}}$ and $D_{\mathrm{upper}}$ are the boundaries of the particle diameter size represented by each of the histograms, each contributing $h_i$.
The result of this fit, shown in Figure~\ref{ref:Fit_spectrum}, is extrapolated to lower \PE\ using the \MC\ simulated samples.

\MC\ simulations of $^{210}$Po $\alpha$-decays from non-coated and TPB-coated dust are used to obtain the corresponding two \FPrompt\ distributions as well as the \R\ distribution integrated over all values of \PIFGAr.
As described under~\Nameref{sec:systematics}, the effective TPB coverage fraction of dust particulates is a floating nuisance parameter in the PLR fit.

\begin{figure}[htb]
 \centering
 \includegraphics[width=1.01\linewidth]{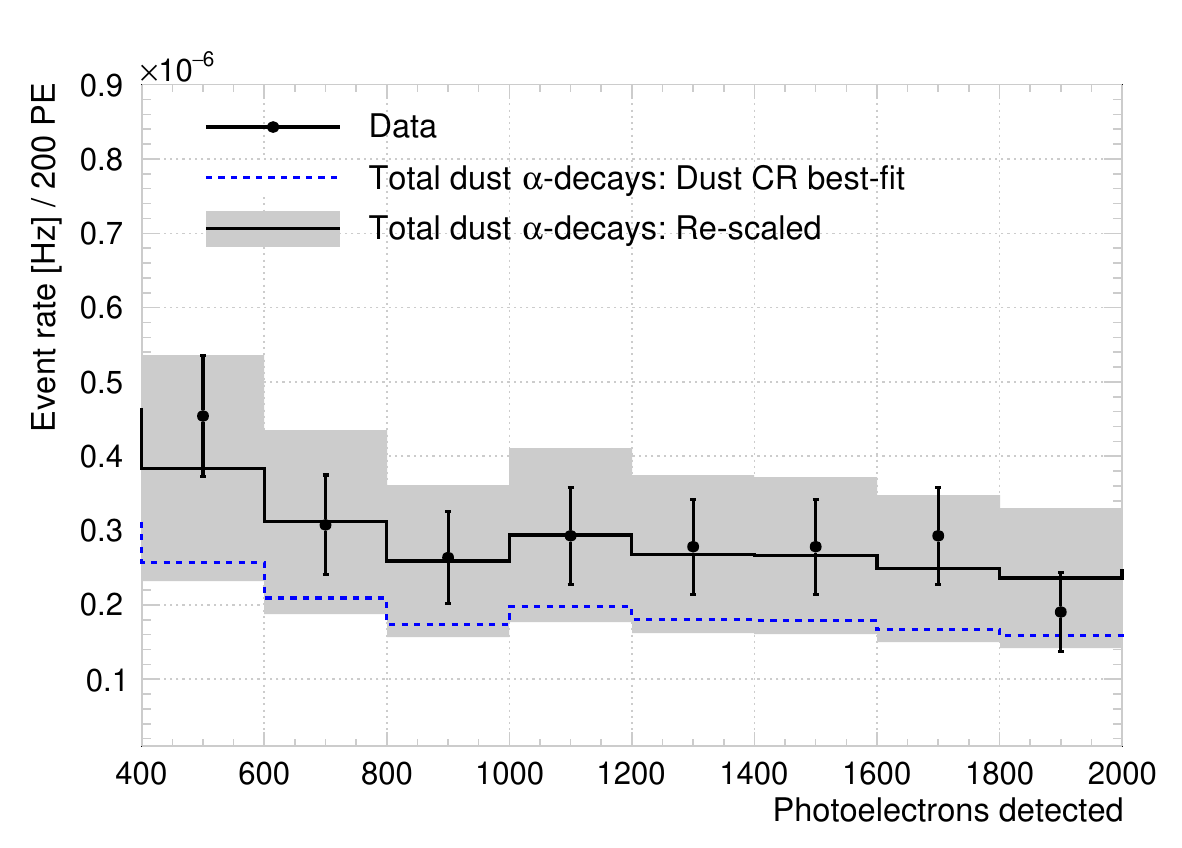}
  \caption{Summed background prediction from dust $\alpha$-decays compared to the data in a sideband of the WIMP ROI.  The prediction is shown before and after applying the normalization correction factor.}
 \label{ref:DustCF}
 \end{figure}

Post-unblinding, we observe an excess of data over the dust $\alpha$-decay background prediction in a sideband of the ROI in the range 400-2000~PE, where no WIMP signal is expected.  
As in the dust CR, the requirement \mbox{$\PIFGAr > 2$} yields a region where dust $\alpha$-decays are the only significant source of events. We use this region to obtain a normalization correction factor for this background, that is applied in the prior of Table~\ref{T:BackgroundPredictionsPost}.  
Figure~\ref{ref:DustCF} compares data and our model in this sideband,
without and with this correction factor of 1.49 applied to the dust $\alpha$-decay background prediction.
This correction factor is applied to the prior background model for dust $\alpha$-decays in the ROI.

\subsubsection*{Surface $\alpha$-Decay Background}\setcurrentname{Surface $\alpha$-Decay Background}\label{subsec:models_surface}

Improving on the surface $\alpha$-decay model from Ref.~\cite{deapdarkmatter2019}, the multi-dimensional PDFs corresponding to this background are now obtained in a data-driven way.
A dedicated ``surface CR'' is defined in the same way as the WIMP ROI, except considering the extended range 90--500~PE and 0.585--0.85 in \FPrompt, without requiring position reconstruction consistency, and requiring \mbox{720 mm $<$ \R\ $<$ 840 mm}.
While the position of most surface $\alpha$-decay events is reconstructed correctly at the radius of the AV inner surface (\R~$ = 850$~mm), events selected in the surface CR are mis-reconstructed surface $\alpha$-decays reconstructing in a shell outside the fiducial volume.
These events are used to obtain the PDFs over the surface CR range in \R, extrapolating inward.

The observed data in the surface CR are expected to contain only surface $\alpha$-decays and dust $\alpha$-decays.  After subtracting the latter contribution using the extrapolation described in the previous section, the surface $\alpha$-decay PDFs in \FPrompt\ and \R\ are obtained from fits to surface CR data over 200--500~PE, and then the one-dimensional PDF describing the PE distribution is obtained over the full range 90--500~PE. As there is no significant difference in the surface $\alpha$-decay PDFs separately for \mbox{$\PIFGAr \le 2$} or \mbox{$\PIFGAr > 2$}, they are taken to be the same in both cases, obtained from the combined dataset observed in the surface CR.

\subsubsection*{WIMP Acceptance}\setcurrentname{WIMP Acceptance}\label{subsec:models_wimp}

\begin{figure}[htb]
 \centering
 \includegraphics[width=1.01\linewidth]{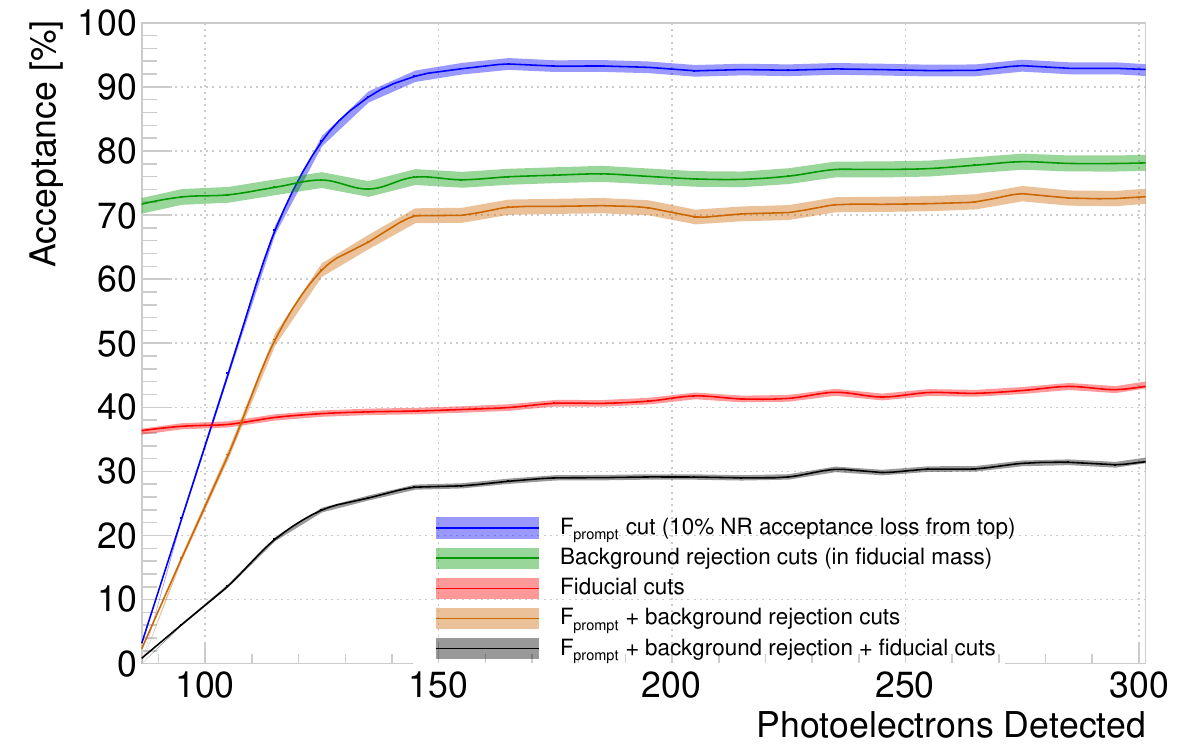}
  \caption{WIMP acceptance as a function of PE under a variety of cut conditions. The total WIMP acceptance is shown in black.
  }
 \label{ref:WIMP_acceptance}
 \end{figure}

Figure \ref{ref:WIMP_acceptance} shows the WIMP acceptance as a function of~PE. ``\FPrompt ~cut'' shows the probability of a WIMP appearing in the \ROI, visible in Figure \ref{F:ROIwithData}. ``Background rejection cuts'' refers to the probability of a WIMP-like event passing the neck veto cut and position reconstruction consistency cuts, given that it has already passed the low-level and fiducial cuts listed in Table \ref{T:Cuts}; the acceptance of these ``fiducial cuts'' is also shown. The total WIMP acceptance, shown in black, is the product of these three curves.

\subsubsection*{PDF Shapes}\setcurrentname{PDF Shapes}\label{subsec:models_dist}

\begin{figure}[t!b]
 \centering
 \includegraphics[width=1.01\linewidth]{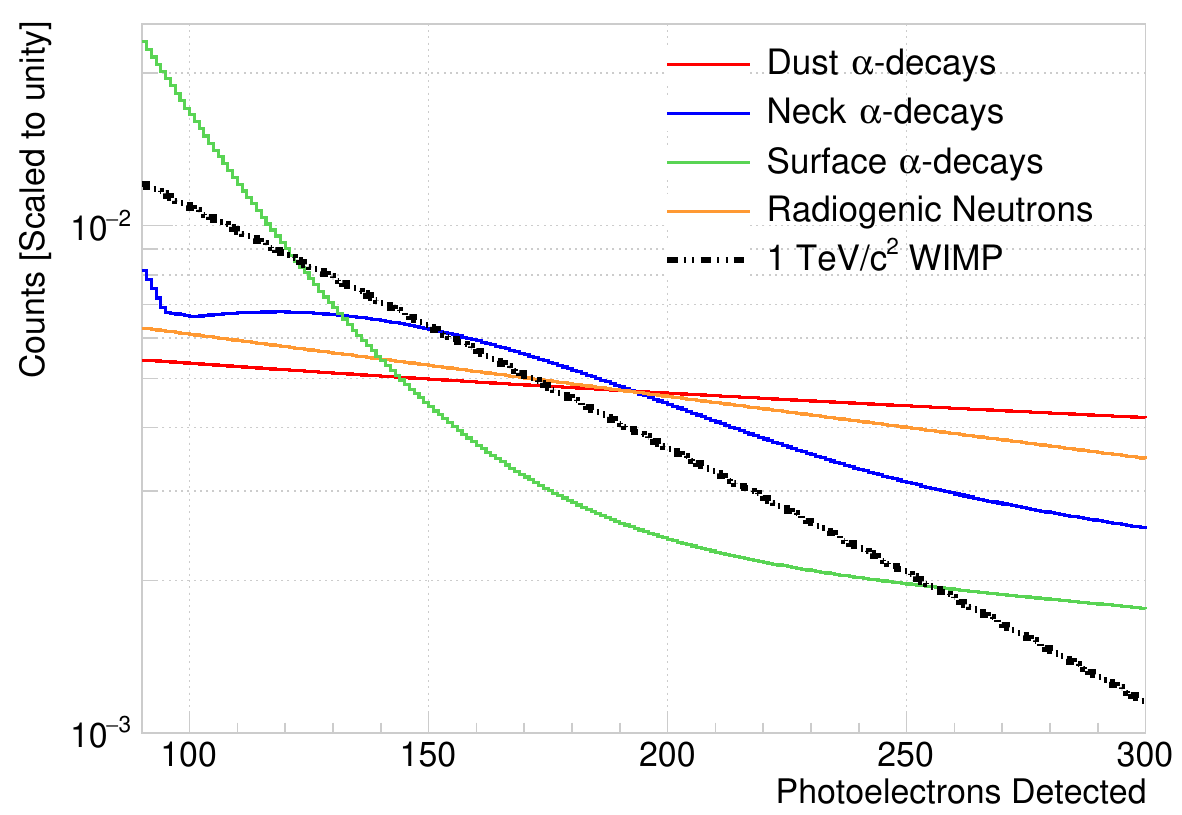}
 \includegraphics[width=1.01\linewidth]{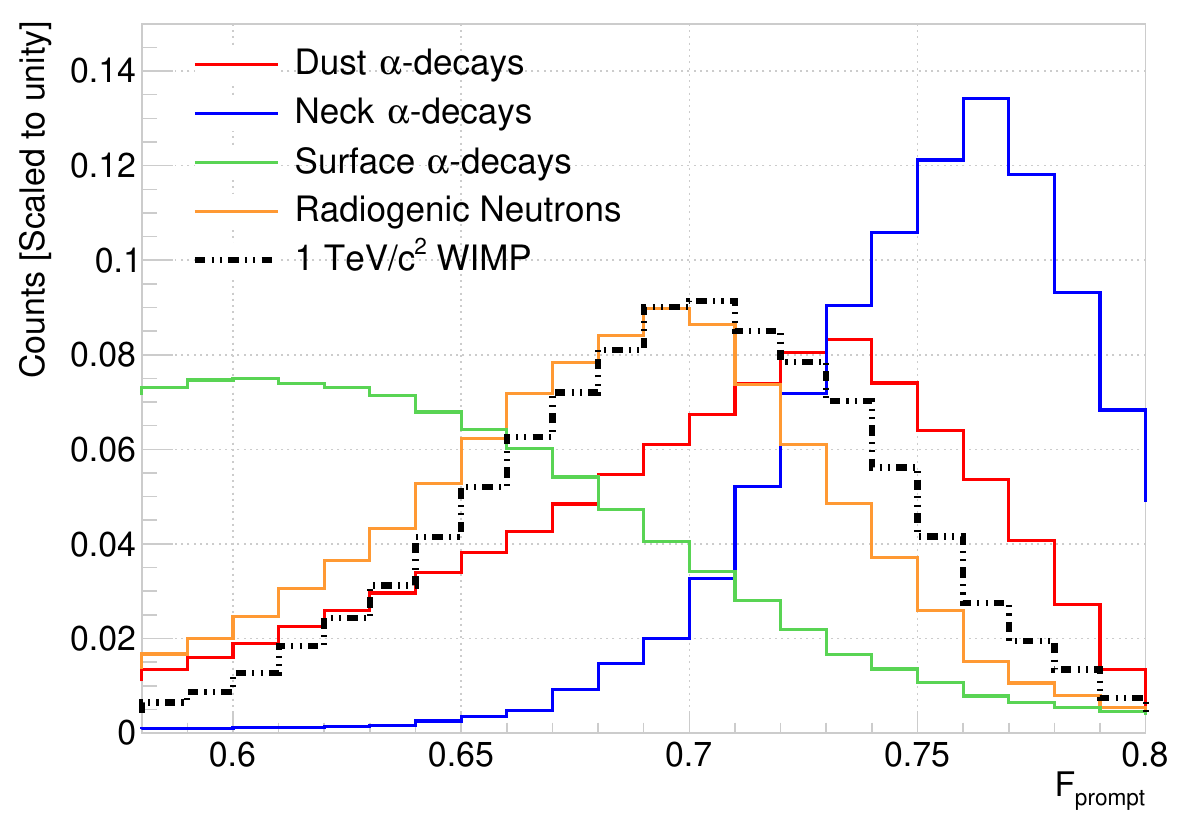}
 \includegraphics[width=1.01\linewidth]{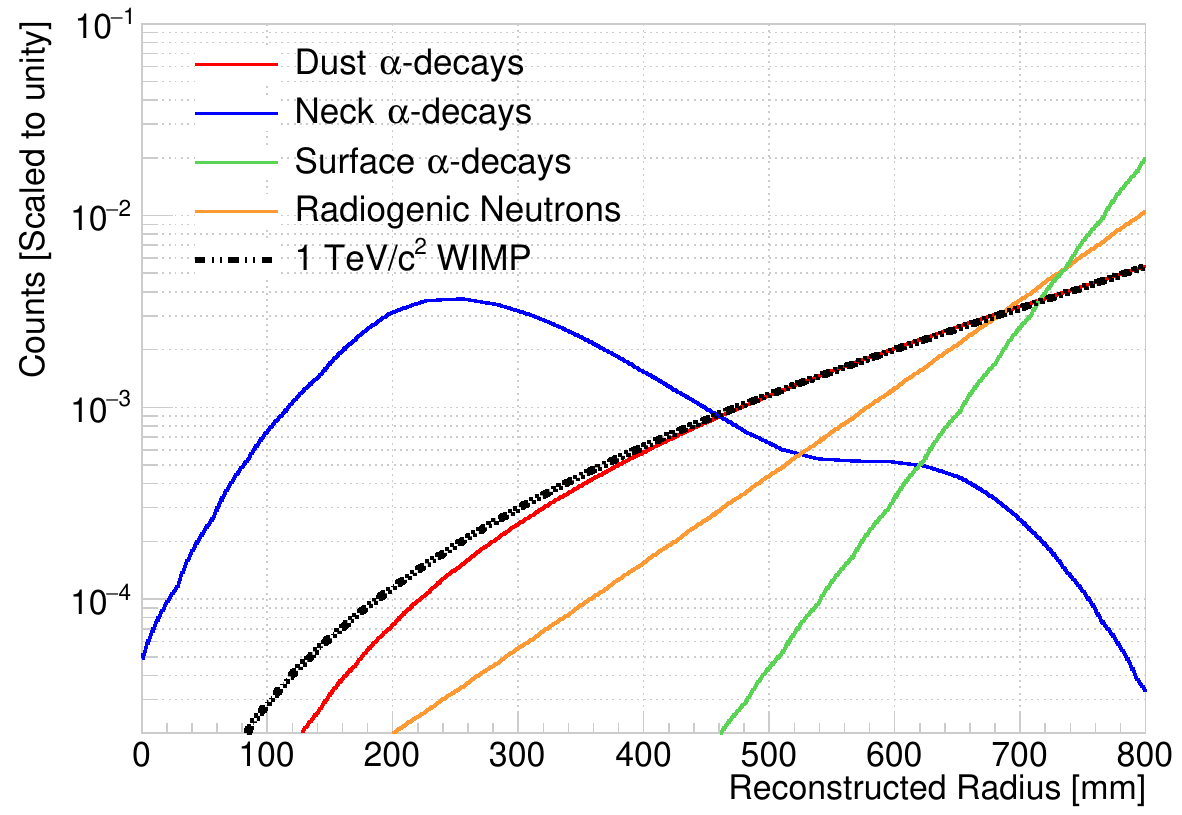}
 \caption{\PE\ (top), \FPrompt\ (middle) and \R\ (bottom) distributions of the five background components implemented in the analysis between \SI{90}{PE}--\SI{300}{PE} (equivalently \SI{14.8}{keV_{ee}}--\SI{49.2}{keV_{ee}}), integrated over all \PIFGAr\ values. Each histogram is normalised to unity for shape comparison. Also shown are the distributions drawn from the signal model for a WIMP mass of 1~TeV/$c^{2}$.
 }
 \label{F:Models_PE_Unity}
\end{figure}

Figure \ref{F:Models_PE_Unity} shows the shapes of the five background model components implemented in the analysis in the three main observable dimensions: \PE, \FPrompt, and \R, integrated over all \PIFGAr\ values. Also shown are the distributions drawn from the signal model, for a WIMP hypothesis with a dark matter particle mass $M_\chi = 1$~TeV/$c^{2}$. 

\subsection*{Systematic Uncertainties}\setcurrentname{Systematic Uncertainties}\label{sec:systematics}

The nuisance parameters used in the \PLR\ implemented here, each corresponding to a specific systematic uncertainty,
are summarized in Table~\ref{T:NuisancePar} and discussed in this section. 

\begin{table}[h!tb]
\begin{center}
\caption{Summary table of the nuisance parameters implemented in the \PLR\ analysis.}
\label{T:NuisancePar}
\begin{tabular}{l}
\hline \hline
\textbf{Systematic Uncertainty (Nuisance Parameter)}            \\ \hline
Light Yield [PE/keV]                  \\ 
Shadowed $\alpha$-decay Light Yield \\
\NR\ Quenching Factor $q_{n}(E_{\mathrm{nr}})$   \\
$\alpha$-Particle Quenching Factor $q_{\alpha}(E_{\mathrm{nr}})$   \\
LAr Optical Model \\
TPB Scattering Length  \\
Dust Particulate TPB Coverage Fraction  \\
Dust $\alpha$-decay $N_{\mathrm{ROI}}$   \\
Shadowed $\alpha$-decay $N_{\mathrm{ROI}}$    \\
Surface $\alpha$-decay $N_{\mathrm{ROI}}$    \\
Radiogenic-n $N_{\mathrm{ROI}}$  \\
\hline \hline
\end{tabular}
\end{center}
\end{table}

\textbf{Light yield.} 
After the calibration described in Section~\Nameref{subsec:reco_charge}, the light yield in \DEAP\ is taken to be \mbox{($6.1 \pm 0.4$)~PE/keV}, where the uncertainty is unchanged from Ref.~\cite{deapdarkmatter2019}.
The light yield nuisance parameter prior is taken to be Gaussian with these mean and width values.

The uncertainty on the light yield of shadowed $\alpha$-decays is larger, accounting for variations on model parameters such as the \LAr\ film thickness and distribution on the flowguides.  Based on comparisons between data and the model in the shadowed $\alpha$-decays CR, this uncertainty is taken to be \mbox{[-50\%, +20\%]}, parametrized using a Gaussian prior with a mean at the nominal light yield and a relative width of 30\%.

\textbf{Quenching factors.} 
The energy-dependent NR quenching $q_{n}$ values and uncertainties are derived from measurements by SCENE \cite{Cao:2015ks},
and validated in \DEAP\ data collected with an AmBe neutron source deployed.

The $\alpha$-particle quenching factor is described as a function of energy based on the product of an electronic quenching factor and a nuclear quenching factor:
\mbox{$q_\alpha = q_{\alpha, {\rm elec}} \times q_{\alpha, {\rm nucl}}$}.
The electronic quenching factor accounts for the non-radiative de-excitation of excimers by interactions with atomic electrons using Birks's formalism~\cite{JBBirks_1951}:
\begin{equation}
q_{\alpha, {\rm elec}}(E_\alpha) = \frac{A}{E_\alpha}\int_{0}^{E_\alpha}{\frac{dE}{1 + B \frac{dE}{dx}}},
\end{equation}
where $A$ accounts for average quenching due to the density of energy deposition within the track core and $B$ accounts for the change in density with energy.
We use the results of the recent \DEAP\ measurement and extrapolation of $q_{\alpha}$ in Ref.~\cite{deapQFalpha}, where $q_{\alpha, {\rm elec}}$ is constrained by considering the \DEAP\ data at the PE peak positions from unattenuated $\alpha$-decays observed from $^{218}$Po and $^{214}$Po in the context of a relative measurement with respect to the value measured by Doke et al.~\cite{DOKE1988291} taken as input, and then extrapolating $q_{\alpha, {\rm elec}}$ to lower $\alpha$-particle energies using the best-fit $A$ and $B$ parameters.

The nuclear quenching factor accounts for the fraction of energy lost by a recoiling nucleus as a result of nuclear collisions.
In Ref.~\cite{deapQFalpha}, the $q_{\alpha, {\rm nucl}}$ calculation is updated using TRIM (TRansport of Ions in Matter) simulations~\cite{SRIM-TRIM}.
To calculate the systematic uncertainty on nuclear quenching, a 10\% relative uncertainty on the energy transferred to electrons and nuclei was applied to the nuclear quenching factor calculated by SRIM (The Stopping and Range of Ions in Matter)-2013~\cite{Ziegler1985} and then scaled by the ratio of TRIM-to-SRIM calculated values.

In this analysis, the overall $\pm 1 \sigma$ uncertainty band on the product $q_{\alpha}$ is implemented as a systematic variation of the PE distributions for $\alpha$-decays.

\textbf{LAr optical model.} 
The optical model of LAr used in \DEAP\ simulations is updated as detailed in Ref.~\cite{westerdale_optics_2023}.  First, the LAr refractive index as a function of wavelength is determined by fitting the Sellmeier equation as a function of wavelength with ultraviolet UV and infrared (IR) poles to measurements by Sinnock and Smith~\cite{sinnock_refractive_1969} and Babicz et al.~\cite{babicz}.  The uncertainties on the fit coefficients are propagated to obtain a wavelength-dependent uncertainty on the refractive index.
The group velocity of light in LAr and the Rayleigh scattering length and their uncertainties are then calculated as a function of wavelength from these LAr refractive index values. In our simulation, one global systematic variation related to the LAr optical model is applied, whereby all three of these LAr optical parameters are varied in a fully correlated way.

\textbf{TPB wavelength shifter.} 
The TPB scattering length is taken to be 2.25~$\mu$m and varied within a factor of two. 
The TPB coverage fraction of dust particulates, bounded between 0 and 1, is varied using a Gaussian prior distribution with a mean of 0.5 and a width of 0.42, consistent with our estimates of systematic uncertainties.

\textbf{Background normalization.} 
The total number of expected ROI events for each background are calculated from the total trigger rates of each component, which are determined using fits to data in CRs defined for this purpose. 
The uncertainties on the background expectation of the four most important components (dust $\alpha$-decays, shadowed $\alpha$-decays, surface $\alpha$-decays, and radiogenic neutrons) are determined using a combination of simulations and in-situ measurements. 

There are no nuisance parameters associated with the ROI expectations for $^{39}$Ar $\beta$-decay, cosmogenic neutron, or Cherenkov backgrounds, as these event types are well-constrained with data from their respective CRs to have a negligible contribution in the WIMP \ROI.

\section*{Data Availability}\setcurrentname{Data Availability}\label{sec:dataavailability}

Data used in this analysis will be made available upon reasonable request to the DEAP collaboration.

\section*{Code Availability}\setcurrentname{Code Availability}\label{sec:codeavailability}

The software used in this paper is a large specialized set of codes for particular computing environments. Therefore, making the software publicly available is impractical. Access to the code may be granted upon request to the DEAP collaboration.

\bibliographystyle{bibliography}
\bibliography{bibliography}

@article{deap_39Ar_specific_activity,
	title = {Precision Measurement of the Specific Activity of 39Ar in Atmospheric Argon with the DEAP-3600 Detector},
	author = {P. Adhikari and others},
	collaboration = {{DEAP-3600 Collaboration}},
	year = {2023},
	journal = {Eur. Phys. J. C},
	volume = {83},
	pages = {642},
	publisher = {{Springer Berlin Heidelberg}},
	doi = {10.1140/epjc/s10052-023-11678-6},
	url = {https://link.springer.com/article/10.1140/epjc/s10052-023-11678-6}
}

@article{deap_39Ar_halflife,
	title = {Direct measurement of the 39Ar half-life from 3.4 years of data with the DEAP-3600 detector},
	author = {P. Adhikari and others},
	collaboration = {{DEAP-3600 Collaboration}},
	year = {2025},
	journal = {Eur. Phys. J. C},
	volume = {85},
	pages = {728},
	publisher = {{Springer Berlin Heidelberg}},
	doi = {10.1140/epjc/s10052-025-14289-5},
	url = {https://doi.org/10.1140/epjc/s10052-025-14289-5}
}

@article{deapdarkmatter2019,
	title = {Search for dark matter with a 231-day exposure of liquid argon using DEAP-3600 at SNOLAB},
	author = {R. Ajaj and P.-A. Amaudruz and G. R. Araujo and M. Baldwin and M. Batygov and B. Beltran and C. E. Bina and J. Bonatt, M. G. Boulay and B. Broerman and J. F. Bueno and P. M. Burghardt and A. Butcher and B. Cai and S. Cavuoti and M. Chen and Y. Chen and B. T. Cleveland and D. Cranshaw and K. Dering and J. DiGioseffo and L. Doria and F. A. Duncan and M. Dunford and A. Erlandson and N. Fatemighomi and G. Fiorillo and S. Florian and A. Flower and R. J. Ford and R. Gagnon and D. Gallacher and E. A. Garc{\'e}s and S. Garg and P. Giampa and D. Goeldi and V. V. Golovko and P. Gorel and K. Graham and D. R. Grant and A. L. Hallin and M. Hamstra and P. J. Harvey and C. Hearns and A. Joy and C. J. Jillings and O. Kamaev and G. Kaur and A. Kemp and I. Kochanek and M. Ku{\'z}niak and S. Langrock and F. La Zia and B. Lehnert and X. Li and J. Lidgard and T. Lindner and O. Litvinov and J. Lock and G. Longo and P. Majewski and A. B. McDonald and T. McElroy and T. McGinn and J. B. McLaughlin and R. Mehdiyev and C. Mielnichuk and J. Monroe and P. Nadeau and C. Nantaisand C. Ng and A. J. Noble and E. O’Dwyer and C. Ouellet and P. Pasuthip and S.J.M. Peeters and M.-C. Piro and T.R. Pollmann and E.T. Rand and C. Rethmeier and F. R{\`e}tiere and N. Seeburn and K. Singhrao and P. Skensved and B. Smith and N. J. T. Smith and T. Sonley and J. Soukup and R. Stainforth and C. Stone and V. Strickland and B. Sur and J. Tang and E. V{\'a}zquez-J{\'a}uregui and L. Veloce and S. Viel and J. Walding and M. Waqar and M. Ward and S. Westerdale and J. Willis and A. Zu{\~n}iga-Reyes},
	collaboration = {{DEAP-3600 Collaboration}},
	journal = {Phys. Rev. D},
	volume = {100},
	number = {2},
	pages = {022004},
	year = {2019},
	month = {Jul},
	publisher = {American Physical Society},
	doi = {10.1103/PhysRevD.100.022004},
	url = {https://link.aps.org/doi/10.1103/PhysRevD.100.022004}
}

@article{deapPosRec,
	title = {Position Reconstruction in the DEAP-3600 Dark Matter Search Experiment},
	author = {P. Adhikari and others},
	collaboration = {{DEAP-3600 Collaboration}},
	year={2025},
	journal={J. Instrum.},
	volume={20},
	pages={P07012},
	url={https://iopscience.iop.org/article/10.1088/1748-0221/20/07/P07012},
}

@article{deapQFalpha,
	title = {Relative Measurement and Extrapolation of the Scintillation Quenching Factor of alpha-Particles in Liquid Argon using DEAP-3600 Data},
	author = {P. Adhikari and others},
	collaboration = {{DEAP-3600 Collaboration}},
	year = {2025},
	journal = {Eur. Phys. J. C},
	volume = {85},
	pages = {87},
	publisher = {{Springer Berlin Heidelberg}},
	doi = {10.1140/epjc/s10052-024-13518-7},
	url = {https://link.springer.com/article/10.1140/epjc/s10052-024-13518-7}
}

@article{DEAP_Pyrene,
title = {Development and characterization of a slow wavelength shifting coating for background rejection in liquid argon detectors},
journal = {Nucl. Instrum. Methods Phys. Res. A},
volume = {1034},
pages = {166683},
year = {2022},
doi = {https://doi.org/10.1016/j.nima.2022.166683},
url = {https://www.sciencedirect.com/science/article/pii/S0168900222002273},
author = {D. Gallacher and others},
}

@article{DEAP_Pyrene_Fluorescence,
doi = {10.1088/1748-0221/16/12/P12029},
url = {https://doi.org/10.1088/1748-0221/16/12/P12029},
year = {2021},
month = {dec},
publisher = {IOP Publishing},
volume = {16},
number = {12},
pages = {P12029},
author = {Benmansour, H. and others},
title = {Fluorescence of pyrene-doped polystyrene films from room temperature down to 4 K for wavelength-shifting applications},
journal = {Journal of Instrumentation},
}

@phdthesis{garg_2023,
	title = {Neck Alpha Background Rejection in {DEAP-3600} using Pyrene and the Search for Superheavy Dark Matter},
	author = {Garg, S.},
	school = {Carleton University},
	address = {Ottawa, Ontario, K1S 5B6, Canada},
	year = {2023}
}

@techreport{hallman1991establishing,
	title={{Establishing a cleanliness program and specifications for the Sudbury Neutrino Observatory}},
	author={Hallman, E. D. and Stokstad, R. G.},
	year={1991},
	number={SNO-STR-91-009},
	publisher={Sudbury Neutrino Observatory}
}

@article{parasuraman2012prediction,
	title={{Prediction model for particle fallout in cleanrooms}},
	author={Parasuraman, Dinesh and Kemps, Antonius and Veeke, Hans and Lodewijks, Gabriel},
	journal={Journal of the IEST},
	volume={55},
	number={1},
	pages={9},
	year={2012},
	publisher={Institute of Environmental Sciences \& Technology},
	doi={10.17764/jiet.55.1.2441h544m128474k},
	url={https://doi.org/10.17764/jiet.55.1.2441h544m128474k}
}

@article{akeribLUXZEPLINLZRadioactivity2020,
	title = {The {{LUX-ZEPLIN}} ({{LZ}}) Radioactivity and Cleanliness Control Programs},
	author = {Akerib, D. S. and Akerlof, C. W. and Akimov, D. Yu and Alquahtani, A. and Alsum, S. K. and Anderson, T. J. and Angelides, N. and Ara{\'u}jo, H. M. and Arbuckle, A. and Armstrong, J. E. and Arthurs, M. and Auyeung, H. and Aviles, S. and Bai, X. and Bailey, A. J. and Balajthy, J. and Balashov, S. and Bang, J. and Barry, M. J. and Bauer, D. and Bauer, P. and Baxter, A. and Belle, J. and Beltrame, P. and Bensinger, J. and Benson, T. and Bernard, E. P. and Bernstein, A. and Bhatti, A. and Biekert, A. and Biesiadzinski, T. P. and Birch, H. J. and Birrittella, B. and Boast, K. E. and Bolozdynya, A. I. and Boulton, E. M. and Boxer, B. and Bramante, R. and Branson, S. and Br{\'a}s, P. and Breidenbach, M. and Brew, C. a. J. and Buckley, J. H. and Bugaev, V. V. and Bunker, R. and Burdin, S. and Busenitz, J. K. and Cabrita, R. and Campbell, J. S. and Carels, C. and Carlsmith, D. L. and Carlson, B. and {Carmona-Benitez}, M. C. and Cascella, M. and Chan, C. and Cherwinka, J. J. and Chiller, A. A. and Chiller, C. and Chott, N. I. and Cole, A. and Coleman, J. and Colling, D. and Conley, R. A. and Cottle, A. and Coughlen, R. and Cox, G. and Craddock, W. W. and Curran, D. and Currie, A. and Cutter, J. E. and da Cunha, J. P. and Dahl, C. E. and Dardin, S. and Dasu, S. and Davis, J. and Davison, T. J. R. and de Viveiros, L. and Decheine, N. and Dobi, A. and Dobson, J. E. Y. and Druszkiewicz, E. and Dushkin, A. and Edberg, T. K. and Edwards, W. R. and Edwards, B. N. and Edwards, J. and Elnimr, M. M. and Emmet, W. T. and Eriksen, S. R. and Faham, C. H. and Fan, A. and Fayer, S. and Fiorucci, S. and Flaecher, H. and Florang, I. M. Fogarty and Ford, P. and Francis, V. B. and Fraser, E. D. and Froborg, F. and Fruth, T. and Gaitskell, R. J. and Gantos, N. J. and Garcia, D. and Gehman, V. M. and Gelfand, R. and Genovesi, J. and Gerhard, R. M. and Ghag, C. and Gibson, E. and Gilchriese, M. G. D. and Gokhale, S. and Gomber, B. and Gonda, T. G. and Greenall, A. and Greenwood, S. and Gregerson, G. and van der Grinten, M. G. D. and Gwilliam, C. B. and Hall, C. R. and Hamilton, D. and Hans, S. and Hanzel, K. and Harrington, T. and Harrison, A. and Harrison, J. and Hasselkus, C. and Haselschwardt, S. J. and Hemer, D. and Hertel, S. A. and Heise, J. and Hillbrand, S. and Hitchcock, O. and Hjemfelt, C. and Hoff, M. D. and Holbrook, B. and Holtom, E. and Hor, J. Y.-K. and Horn, M. and Huang, D. Q. and Hurteau, T. W. and Ignarra, C. M. and Irving, M. N. and Jacobsen, R. G. and Jahangir, O. and Jeffery, S. N. and Ji, W. and Johnson, M. and Johnson, J. and Johnson, P. and Jones, W. G. and Kaboth, A. C. and Kamaha, A. and Kamdin, K. and Kasey, V. and Kazkaz, K. and Keefner, J. and Khaitan, D. and Khaleeq, M. and Khazov, A. and Khromov, A. V. and Khurana, I. and Kim, Y. D. and Kim, W. T. and Kocher, C. D. and Kodroff, D. and Konovalov, A. M. and Korley, L. and Korolkova, E. V. and Koyuncu, M. and Kras, J. and Kraus, H. and Kravitz, S. W. and Krebs, H. J. and Kreczko, L. and Krikler, B. and Kudryavtsev, V. A. and Kumpan, A. V. and Kyre, S. and Lambert, A. R. and Landerud, B. and Larsen, N. A. and Laundrie, A. and Leason, E. A. and Lee, H. S. and Lee, J. and Lee, C. and Lenardo, B. G. and Leonard, D. S. and Leonard, R. and Lesko, K. T. and Levy, C. and Li, J. and Liu, Y. and Liao, J. and Liao, F.-T. and Lin, J. and Lindote, A. and Linehan, R. and Lippincott, W. H. and Liu, R. and Liu, X. and Loniewski, C. and Lopes, M. I. and {Lopez-Asamar}, E. and Paredes, B. L{\'o}pez and Lorenzon, W. and Lucero, D. and Luitz, S. and Lyle, J. M. and Lynch, C. and Majewski, P. A. and Makkinje, J. and Malling, D. C. and Manalaysay, A. and Manenti, L. and Mannino, R. L. and Marangou, N. and Markley, D. J. and MarrLaundrie, P. and Martin, T. J. and Marzioni, M. F. and Maupin, C. and McConnell, C. T. and McKinsey, D. N. and McLaughlin, J. and Mei, D.-M. and Meng, Y. and Miller, E. H. and Minaker, Z. J. and Mizrachi, E. and Mock, J. and Molash, D. and Monte, A. and Monzani, M. E. and Morad, J. A. and Morrison, E. and Mount, B. J. and Murphy, A. St J. and Naim, D. and Naylor, A. and Nedlik, C. and Nehrkorn, C. and Nelson, H. N. and Nesbit, J. and Neves, F. and Nikkel, J. A. and Nikoleyczik, J. A. and Nilima, A. and O'Dell, J. and Oh, H. and O'Neill, F. G. and O'Sullivan, K. and Olcina, I. and Olevitch, M. A. and {Oliver-Mallory}, K. C. and Oxborough, L. and Pagac, A. and Pagenkopf, D. and Pal, S. and Palladino, K. J. and Palmaccio, V. M. and Palmer, J. and Pangilinan, M. and Parveen, N. and Patton, S. J. and Pease, E. K. and Penning, B. P. and Pereira, G. and Pereira, C. and Peterson, I. B. and Piepke, A. and Pierson, S. and Powell, S. and Preece, R. M. and Pushkin, K. and Qie, Y. and Racine, M. and Ratcliff, B. N. and Reichenbacher, J. and Reichhart, L. and Rhyne, C. A. and Richards, A. and Riffard, Q. and Rischbieter, G. R. C. and Rodrigues, J. P. and Rose, H. J. and Rosero, R. and Rossiter, P. and Rucinski, R. and Rutherford, G. and Saba, J. S. and Sabarots, L. and Santone, D. and Sarychev, M. and Sazzad, A. B. M. R. and Schnee, R. W. and Schubnell, M. and Scovell, P. R. and Severson, M. and Seymour, D. and Shaw, S. and Shutt, G. W. and Shutt, T. A. and Silk, J. J. and Silva, C. and Skarpaas, K. and Skulski, W. and Smith, A. R. and Smith, R. J. and Smith, R. E. and So, J. and Solmaz, M. and Solovov, V. N. and Sorensen, P. and Sosnovtsev, V. V. and Stancu, I. and Stark, M. R. and Stephenson, S. and Stern, N. and Stevens, A. and Stiegler, T. M. and Stifter, K. and Studley, R. and Sumner, T. J. and Sundarnath, K. and Sutcliffe, P. and Swanson, N. and Szydagis, M. and Tan, M. and Taylor, W. C. and Taylor, R. and Taylor, D. J. and Temples, D. and Tennyson, B. P. and Terman, P. A. and Thomas, K. J. and Thomson, J. A. and Tiedt, D. R. and Timalsina, M. and To, W. H. and Tom{\'a}s, A. and Tope, T. E. and Tripathi, M. and Tronstad, D. R. and Tull, C. E. and Turner, W. and Tvrznikova, L. and Utes, M. and Utku, U. and Uvarov, S. and Va'vra, J. and Vacheret, A. and Vaitkus, A. and Verbus, J. R. and Vietanen, T. and Voirin, E. and Vuosalo, C. O. and Walcott, S. and Waldron, W. L. and Walker, K. and Wang, J. J. and Wang, R. and Wang, L. and Wang, W. and Wang, Y. and Watson, J. R. and Migneault, J. and Weatherly, S. and Webb, R. C. and Wei, W.-Z. and While, M. and White, R. G. and White, J. T. and White, D. T. and Whitis, T. J. and Wisniewski, W. J. and Wilson, K. and Witherell, M. S. and Wolfs, F. L. H. and Wolfs, J. D. and Woodward, D. and Worm, S. D. and Xiang, X. and Xiao, Q. and Xu, J. and Yeh, M. and Yin, J. and Young, I. and Zhang, C. and Zarzhitsky, P.},
	year = {2020},
	month = nov,
	journal = {Eur. Phys. J. C},
	volume = {80},
	number = {11},
	pages = {1044},
	publisher = {{Springer Berlin Heidelberg}},
	issn = {1434-6044, 1434-6052},
	doi = {10.1140/epjc/s10052-020-8420-x},
	url = {https://epjc.epj.org/articles/epjc/abs/2020/11/10052_2020_Article_8420/10052_2020_Article_8420.html}
}

@article{amaudruz_design_2017,
	title = {{Design and construction of the DEAP-3600 dark matter detector}},
	journal = {Astropart. Phys.},
	volume = {108},
	pages = {1--23},
	year = {2019},
	doi = {10.1016/j.astropartphys.2018.09.006},
	url = {http://www.sciencedirect.com/science/article/pii/S0927650518300914},
	author = {P.-A. Amaudruz and M. Baldwin and M. Batygov and B. Beltran and C.E. Bina and D. Bishop and J. Bonatt and G. Boorman and M.G. Boulay and B. Broerman and T. Bromwich and J.F. Bueno and P.M. Burghardt and A. Butcher and B. Cai and S. Chan and M. Chen and R. Chouinard and S. Churchwell and B.T. Cleveland and D. Cranshaw and K. Dering and J. DiGioseffo and S. Dittmeier and F.A. Duncan and M. Dunford and A. Erlandson and N. Fatemighomi and S. Florian and A. Flower and R.J. Ford and R. Gagnon and P. Giampa and V.V. Golovko and P. Gorel and R. Gornea and E. Grace and K. Graham and D.R. Grant and E. Gulyev and A. Hall and A.L. Hallin and M. Hamstra and P.J. Harvey and C. Hearns and C.J. Jillings and O. Kamaev and A. Kemp and M. Ku{\'z}niak and S. Langrock and F. La Zia and B. Lehnert and O. Li and J.J. Lidgard and P. Liimatainen and C. Lim and T. Lindner and Y. Linn and S. Liu and P. Majewski and R. Mathew and A.B. McDonald and T. McElroy and K. McFarlane and T. McGinn and J.B. McLaughlin and S. Mead and R. Mehdiyev and C. Mielnichuk and J. Monroe and A. Muir and P. Nadeau and C. Nantais and C. Ng and A.J. Noble and E. O’Dwyer and C. Ohlmann and K. Olchanski and K.S. Olsen and C. Ouellet and P. Pasuthip and S.J.M. Peeters and T.R. Pollmann and E.T. Rand and W. Rau and C. Rethmeier and F. Reti{\`e}re and N. Seeburn and B. Shaw and K. Singhrao and P. Skensved and B. Smith and N.J.T. Smith and T. Sonley and J. Soukup and R. Stainforth and C. Stone and V. Strickland and B. Sur and J. Tang and J. Taylor and L. Veloce and E. V{\'a}zquez-J{\'a}uregui and J. Walding and M. Ward and S. Westerdale and R. White and E. Woolsey and J. Zielinski},
	collaboration = {{DEAP-3600 Collaboration}},
}

@article{Cao:2015ks,
	title = {{Measurement of scintillation and ionization yield and scintillation pulse shape from nuclear recoils in liquid argon}},
	author = {Cao, H. and Alexander, T. and Aprahamian, A. and Avetisyan, R. and Back, H. O. and Cocco, A. G. and DeJongh, F. and Fiorillo, G. and Galbiati, C. and Grandi, L. and Guardincerri, Y. and Kendziora, C. and Lippincott, W. H. and Love, C. and Lyons, S. and Manenti, L. and Martoff, C. J. and Meng, Y. and Montanari, D. and Mosteiro, P. and Olvitt, D. and Pordes, S. and Qian, H. and Rossi, B. and Saldanha, R. and Sangiorgio, S. and Siegl, K. and Strauss, S. Y. and Tan, W. and Tatarowicz, J. and Walker, S. and Wang, H. and Watson, A. W. and Westerdale, S. and Yoo, J.},
	collaboration = {{SCENE Collaboration}},
	journal = {Phys. Rev. D},
	volume = {91},
	issue = {9},
	pages = {092007},
	numpages = {29},
	year = {2015},
	month = {May},
	publisher = {American Physical Society},
	doi = {10.1103/PhysRevD.91.092007},
	url = {https://link.aps.org/doi/10.1103/PhysRevD.91.092007}
}

@article{westerdale_optics_2023,
author = {Westerdale, Shawn},
title = {{The DEAP-3600 liquid argon optical model and NEST updates}},
collaboration = {{on behalf of DEAP-3600 and NEST Collaborations}},
journal = {arXiv},
year = {2023},
eprint = {2312.07712},
eprinttype = {arxiv},
eprintclass = {physics.ins-det},
month = dec,
url = {http://arxiv.org/abs/2312.07712},
}

@article{the_deap_collaboration_-situ_2017,
	title = {{In-situ characterization of the Hamamatsu R5912-HQE photomultiplier tubes used in the DEAP-3600 experiment}},
	journal = {Nucl. Instrum. Methods Phys. Res. A},
	volume = {922},
	pages = {373--384},
	year = {2019},
	doi = {10.1016/j.nima.2018.12.058},
	url = {http://www.sciencedirect.com/science/article/pii/S0168900218318552},
	author = {P.-A. Amaudruz and M. Batygov and B. Beltran and C.E. Bina and D. Bishop and J. Bonatt and G. Boorman and M.G. Boulay and B. Broerman and T. Bromwich and J.F. Bueno and A. Butcher and B. Cai and S. Chan and M. Chen and R. Chouinard and S. Churchwell and B.T. Cleveland and D. Cranshaw and K. Dering and S. Dittmeier and F.A. Duncan and M. Dunford and A. Erlandson and N. Fatemighomi and R.J. Ford and R. Gagnon and P. Giampa and V.V. Golovko and P. Gorel and R. Gornea and E. Grace and K. Graham and D.R. Grant and E. Gulyev and A. Hall and A.L. Hallin and M. Hamstra and P.J. Harvey and C. Hearns and C.J. Jillings and O. Kamaev and A. Kemp and M. Ku{\'z}niak and S. Langrock and F. La Zia and B. Lehnert and O. Li and J.J. Lidgard and P. Liimatainen and C. Lim and T. Lindner and Y. Linn and S. Liu and R. Mathew and A.B. McDonald and T. McElroy and K. McFarlane and J. McLaughlin and S. Mead and R. Mehdiyev and C. Mielnichuk and J. Monroe and A. Muir and P. Nadeau and C. Nantais and C. Ng and A.J. Noble and E. O’Dwyer and C. Ohlmann and K. Olchanski and K.S. Olsen and C. Ouellet and P. Pasuthip and S.J.M. Peeters and T.R. Pollmann and E.T. Rand and W. Rau and C. Rethmeier and F. Reti{\`e}re and N. Seeburn and B. Shaw and K. Singhrao and P. Skensved and B. Smith and N.J.T. Smith and T. Sonley and R. Stainforth and C. Stone and V. Strickland and B. Sur and J. Tang and J. Taylor and L. Veloce and E. V{\'a}zquez-J{\'a}uregui and J. Walding and M. Ward and S. Westerdale and R. White and E. Woolsey and J. Zielinski},
	collaboration = {{DEAP-3600 Collaboration}}
}

@article{lindner_deap-3600_2015,
	author={Lindner, Thomas},
	title={{DEAP-3600 Data Acquisition System}},
	journal={J. Phys. Conf. Ser.},
	volume={664},
	number={8},
	pages={082026},
	url={http://stacks.iop.org/1742-6596/664/i=8/a=082026},
	doi = {10.1088/1742-6596/664/8/082026},
	year={2015}
}

@misc{rat_github,
	author = {Bolton, T. and Gastler, D. and Klein, J. and Lippincott, H. and Mastbaum, A. and Nikkel, J. and Orebi Gann, G. and Akashi-Ronquest, M. and Seibert, S. and Sekula, S. and Seligman, W. and Tunnell, C. and Worcester, M.},
	title = {{RAT (is an Analysis Tool) User's Guide}},
	year = {2018},
	url = {https://rat.readthedocs.io}
}

@phdthesis{butcher_2015,
	title = {{Searching for dark matter with DEAP-3600}},
	author = {Butcher, A.},
	school = {Royal Holloway, University of London},
	address = {Egham, Surrey, TW20 0EX, UK},
	year = {2015}
}

@mastersthesis{burghardt_2018,
	title = {{Background suppression through pulse shape analysis in the DEAP-3600 dark matter detector}},
	author = {Burghart, M.},
	school = {Technical University of Munich},
	address = {Arcisstrasse 21, 80333 M{\"u}nchen, Germany},
	year = {2018}
}

@mastersthesis{delgobbo_2021,
	title = {{Attenuated alpha backgrounds in the DEAP-3600 dark matter search experiment}},
	author = {DelGobbo, P.},
	school = {Carleton University},
	address = {1125 Colonel By Drive, Ottawa, ON, K1S 5B6, Canada},
	year = {2021}
}

@article{planck_collaboration_planck_2018,
	author = {Aghanim, N. and Akrami, Y. and Ashdown, M. and Aumont, J. and Baccigalupi, C. and Ballardini, M. and Banday, A. J. and Barreiro, R. B. and Bartolo, N. and Basak, S. and Battye, R. and Benabed, K. and Bernard, J.-P. and Bersanelli, M. and Bielewicz, P. and Bock, J. J. and Bond, J. R. and Borrill, J. and Bouchet, F. R. and Boulanger, F. and Bucher, M. and Burigana, C. and Butler, R. C. and Calabrese, E. and Cardoso, J.-F. and Carron, J. and Challinor, A. and Chiang, H. C. and Chluba, J. and Colombo, L. P. L. and Combet, C. and Contreras, D. and Crill, B. P. and Cuttaia, F. and de Bernardis, P. and de Zotti, G. and Delabrouille, J. and Delouis, J.-M. and Di Valentino, E. and Diego, J. M. and Doré, O. and Douspis, M. and Ducout, A. and Dupac, X. and Dusini, S. and Efstathiou, G. and Elsner, F. and Enßlin, T. A. and Eriksen, H. K. and Fantaye, Y. and Farhang, M. and Fergusson, J. and Fernandez-Cobos, R. and Finelli, F. and Forastieri, F. and Frailis, M. and Franceschi, E. and Frolov, A. and Galeotta, S. and Galli, S. and Ganga, K. and Génova-Santos, R. T. and Gerbino, M. and Ghosh, T. and González-Nuevo, J. and Górski, K. M. and Gratton, S. and Gruppuso, A. and Gudmundsson, J. E. and Hamann, J. and Handley, W. and Herranz, D. and Hivon, E. and Huang, Z. and Jaffe, A. H. and Jones, W. C. and Karakci, A. and Keihänen, E. and Keskitalo, R. and Kiiveri, K. and Kim, J. and Kisner, T. S. and Knox, L. and Krachmalnicoff, N. and Kunz, M. and Kurki-Suonio, H. and Lagache, G. and Lamarre, J.-M. and Lasenby, A. and Lattanzi, M. and Lawrence, C. R. and Jeune, M. Le and Lemos, P. and Lesgourgues, J. and Levrier, F. and Lewis, A. and Liguori, M. and Lilje, P. B. and Lilley, M. and Lindholm, V. and López-Caniego, M. and Lubin, P. M. and Ma, Y.-Z. and Macías-Pérez, J. F. and Maggio, G. and Maino, D. and Mandolesi, N. and Mangilli, A. and Marcos-Caballero, A. and Maris, M. and Martin, P. G. and Martinelli, M. and Martínez-González, E. and Matarrese, S. and Mauri, N. and McEwen, J. D. and Meinhold, P. R. and Melchiorri, A. and Mennella, A. and Migliaccio, M. and Millea, M. and Mitra, S. and Miville-Deschênes, M.-A. and Molinari, D. and Montier, L. and Morgante, G. and Moss, A. and Natoli, P. and Nørgaard-Nielsen, H. U. and Pagano, L. and Paoletti, D. and Partridge, B. and Patanchon, G. and Peiris, H. V. and Perrotta, F. and Pettorino, V. and Piacentini, F. and Polastri, L. and Polenta, G. and Puget, J.-L. and Rachen, J. P. and Reinecke, M. and Remazeilles, M. and Renzi, A. and Rocha, G. and Rosset, C. and Roudier, G. and Rubiño-Martín, J. A. and Ruiz-Granados, B. and Salvati, L. and Sandri, M. and Savelainen, M. and Scott, D. and Shellard, E. P. S. and Sirignano, C. and Sirri, G. and Spencer, L. D. and Sunyaev, R. and Suur-Uski, A.-S. and Tauber, J. A. and Tavagnacco, D. and Tenti, M. and Toffolatti, L. and Tomasi, M. and Trombetti, T. and Valenziano, L. and Valiviita, J. and Van Tent, B. and Vibert, L. and Vielva, P. and Villa, F. and Vittorio, N. and Wandelt, B. D. and Wehus, I. K. and White, M. and White, S. D. M. and Zacchei, A. and Zonca, A.},
	title = {{Planck 2018 results. VI. Cosmological parameters}},
	collaboration = {{Planck Collaboration}},
	year = {2020},
	journal = {Astronomy and Astrophysics},
	volume = {641},
	pages = {{A6}},
	url = {http://arxiv.org/abs/1807.06209}
}

@article{brun_root_1997,
	title = {{ROOT---An object oriented data analysis framework}},
	journal = {Nucl. Instrum. Methods Phys. Res. A},
	volume = {389},
	number = {1},
	pages = {81--86},
	year = {1997},
	note = {{New Computing Techniques in Physics Research V}},
	doi = {10.1016/S0168-9002(97)00048-X},
	url = {http://www.sciencedirect.com/science/article/pii/S016890029700048X},
	author = {Brun, Rene and Rademakers, Fons}
}

@article{mccabe_astrophysical_2010,
	title = {{Astrophysical uncertainties of dark matter direct detection experiments}},
	author = {McCabe, Christopher},
	journal = {Phys. Rev. D},
	volume = {82},
	issue = {2},
	pages = {023530},
	numpages = {13},
	year = {2010},
	month = {Jul},
	publisher = {American Physical Society},
	doi = {10.1103/PhysRevD.82.023530},
	url = {https://link.aps.org/doi/10.1103/PhysRevD.82.023530}
}

@article{Agostinelli:2003fg,
	title = {{Geant4---a simulation toolkit}},
	journal = {Nucl. Instrum. Methods Phys. Res. A},
	volume = {506},
	number = {3},
	pages = {250--303},
	year = {2003},
	doi = {10.1016/S0168-9002(03)01368-8},
	url = {http://www.sciencedirect.com/science/article/pii/S0168900203013688},
	author = {Agostinelli, S and Allison, J and Amako, K and Apostolakis, J and Araujo, H and Arce, P and Asai, M and Axen, D and Banerjee, S and Barrand, G and Behner, F and Bellagamba, L and Boudreau, J and Broglia, L and Brunengo, A and Burkhardt, H and Chauvie, S and Chuma, J and Chytracek, R and Cooperman, G and Cosmo, G and Degtyarenko, P and Dell{\textquoteright}Acqua, A and Depaola, G and Dietrich, D and Enami, R and Feliciello, A and Ferguson, C and Fesefeldt, H and Folger, G and Foppiano, F and Forti, A and Garelli, S and Giani, S and Giannitrapani, R and Gibin, D and G{\'o}mez-Cadenas, J J and Gonz{\'a}lez, I and Gracia Abril, G and Greeniaus, G and Greiner, W and Grichine, V and Grossheim, A and Guatelli, S and Gumplinger, P and Hamatsu, R and Hashimoto, K and Hasui, H and Heikkinen, A and Howard, A and Ivanchenko, V and Johnson, A and Jones, F W and Kallenbach, J and Kanaya, N and Kawabata, M and Kawabata, Y and Kawaguti, M and Kelner, S and Kent, P and Kimura, A and Kodama, T and Kokoulin, R and Kossov, M and Kurashige, H and Lamanna, E and Lamp{\'e}n, T and Lara, V and Lefebure, V and Lei, F and Liendl, M and Lockman, W and Longo, F and Magni, Stefano and Maire, M and Medernach, E and Minamimoto, K and Mora de Freitas, P and Morita, Y and Murakami, K and Nagamatu, M and Nartallo, R and Nieminen, P and Nishimura, T and Ohtsubo, K and Okamura, M and O'Neale, S and Oohata, Y and Paech, K and Perl, J and Pfeiffer, A and Pia, M G and Ranjard, F and Rybin, A and Sadilov, S and Di Salvo, E and Santin, G and Sasaki, T and Savvas, N and Sawada, Y and Scherer, S and Sei, S and Sirotenko, V and Smith, D and Starkov, N and Stoecker, H and Sulkimo, J and Takahata, M and Tanaka, S and Tcherniaev, E and Safai Tehrani, E and Tropeano, M and Truscott, P and Uno, H and Urban, L and Urban, P and Verderi, M and Walkden, A and Wander, W and Weber, H and Wellisch, J P and Wenaus, T and Williams, D C and Wright, D and Yamada, T and Yoshida, H and Zschiesche, D}
}

@article{sinnock_refractive_1969,
	title = {{Refractive indices of the condensed inert gases}},
	author = {Sinnock, A. C. and Smith, B. L.},
	journal = {Phys. Rev.},
	volume = {181},
	issue = {3},
	pages = {1297--1307},
	numpages = {0},
	year = {1969},
	month = {May},
	publisher = {American Physical Society},
	doi = {10.1103/PhysRev.181.1297},
	url = {https://link.aps.org/doi/10.1103/PhysRev.181.1297}
}

@article{DOKE1988291,
title = {LET dependence of scintillation yields in liquid argon},
journal = {Nucl. Instrum. Methods Phys. Res. A},
volume = {269},
number = {1},
pages = {291-296},
year = {1988},
doi = {https://doi.org/10.1016/0168-9002(88)90892-3},
url = {https://www.sciencedirect.com/science/article/pii/0168900288908923},
author = {Tadayoshi Doke and Henry J. Crawford and Akira Hitachi and Jun Kikuchi and Peter J. Lindstrom and Kimiaki Masuda and Eido Shibamura and Tan Takahashi}
}

@article{cowan2011asymptotic,
	title={Asymptotic formulae for likelihood-based tests of new physics},	
 	author={Cowan, Glen and Cranmer, Kyle and Gross, Eilam and Vitells, Ofer},
	journal = {Eur. Phys. J. C},
	year = {2011},
	volume = {71},
	number = {2},
	pages={1--19},
	doi={test}
}

@article{statswhitepaper,
	title={Recommended conventions for reporting results from direct dark matter searches},	
 	author={Baxter, D and Bloch, IM and Bodnia, E and Chen, X and Conrad, Jan and Di Gangi, P and Dobson, JEY and Durnford, D and Haselschwardt, SJ and Kaboth, A and others},
	journal = {Eur. Phys. J. C},
	year = {2021},
	volume = {81},
	number = {10},
	pages={1--19},
	doi={test}
}

@article{DEAP_PSD,
	title={Pulseshape discrimination against low-energy Ar-39 beta decays in liquid argon with 4.5 tonne-years of DEAP-3600 data},
	collaboration = {{DEAP-3600 Collaboration}},
	author= {Adhikari, P and others},
	journal = {Eur. Phys. J. C},
	volume = {81},
	pages = {823},
	year = {2021},
	doi = {10.1140/epjc/s10052-021-09514-w},
	url = {https://doi.org/10.1140/epjc/s10052-021-09514-w}
}

@article{minuit,
	title={Developments of mathematical software libraries for the LHC experiments},
	author= {Hatlo, M and James, F and Mato, P and Moneta, L and Winkler, M and Zsenei, A},
	journal = {IEEE Transactions of Nuclear Science},
	volume = {52},
	number = {6},
	year = {2005}, 
	doi={test}
}

@article{babicz,
	title = {{Experimental study of the propagation of scintillation light in Liquid Argon}},
	author = {Babicz, M and Bordoni, S and Cervi, T and Collins, Z and Fava, A and Kose, U and Meli, M and Menegolli, A and Nessi, M and Pietropaolo, F and others},
	journal = {Nucl. Instrum. Methods Phys. Res. A},
	volume = {936},
	pages = {178--179},
	year = {2019},
	doi = {test}
}

@article{JBBirks_1951,
	title = {Scintillations from Organic Crystals: Specific Fluorescence and Relative Response to Different Radiations},	
	author = {J B Birks},
	journal = {Proceedings of the Physical Society. Section A},
	volume = {64},	
	number = {10},
	pages = {874},
	year = {1951},
	doi = {10.1088/0370-1298/64/10/303},
	url = {https://dx.doi.org/10.1088/0370-1298/64/10/303}
}

@misc{SRIM-TRIM,
author="Ziegler, James F.
and Biersack, Jochen P. and Ziegler, Matthias D. ",
title="{SRIM} software",
url="http://www.srim.org/"
}

@misc{Ziegler1985,
author="Ziegler, James F.
and Biersack, Jochen P.",
title="{The Stopping and Range of Ions in Matter}",
bookTitle="Treatise on Heavy-Ion Science: Volume 6: Astrophysics, Chemistry, and Condensed Matter",
year="1985",
publisher="Springer US",
address="Boston, MA",
pages="93--129",
isbn = "978-1-4615-8103-1",
doi="10.1007/978-1-4615-8103-1_3",
url="https://doi.org/10.1007/978-1-4615-8103-1_3"
}

\section*{Acknowledgments}
We thank the Natural Sciences and Engineering Research Council of Canada (NSERC),
the Canada Foundation for Innovation (CFI),
the Ontario Ministry of Research and Innovation (MRI), 
and Alberta Advanced Education and Technology (ASRIP),
the University of Alberta,
Carleton University, 
Queen's University,
the Canada First Research Excellence Fund through the Arthur B.~McDonald Canadian Astroparticle Physics Research Institute,
SECIHTI Project No. CBF-2025-I-1589,
DGAPA UNAM Grants No. PAPIIT IN105923 and IN102326,
the European Research Council Project (ERC StG 279980),
the UK Science and Technology Facilities Council (STFC) (ST/K002570/1 and ST/R002908/1),
the Leverhulme Trust (ECF-20130496),
the Russian Science Foundation (Grant No. 21-72-10065),
the Spanish Ministry of Science and Innovation (PID2019-109374GB-I00) and the Community of Madrid (2018-T2/ TIC-10494), 
the Polish National Science Centre (2022/47/B/ST2/02015),
the European Union,
International Research Agenda Programmes of the Foundation for Polish Science: AstroCeNT (MAB/2018/7), funded from the European Regional Development Fund, and Astrocent, co-financed by the European Union under the European Funds for Smart Economy 2021-2027 (FENG).
Studentship support from
the Rutherford Appleton Laboratory Particle Physics Division,
STFC and SEPNet PhD is acknowledged.
We thank SNOLAB and its staff for support through underground space, logistical, and technical services.
SNOLAB operations are supported by the CFI
and Province of Ontario MRI,
with underground access provided by Vale at the Creighton mine site.
We thank Vale for their continuing support, including the work of shipping the acrylic vessel underground.
We gratefully acknowledge the support of the Digital Research Alliance of Canada,
Calcul Qu\'ebec,
the Centre for Advanced Computing at Queen's University,
and the Computational Centre for Particle and Astrophysics (C2PAP) at the Leibniz Supercomputer Centre (LRZ)
for providing the computing resources required to undertake this work.

\section*{Author Contributions}\setcurrentname{Author Contributions}\label{sec:authorcont}

All authors have contributed to this publication, being variously involved in the detector design, construction, operation, data taking and data analysis, discussing and approving the scientific results. This article was prepared by a subgroup of authors from the DEAP collaboration and subjected to an internal collaboration-wide review process. 

\section*{Competing Interests}\setcurrentname{Competing Interests}\label{sec:competinginterests}

The authors declare no competing interests.

\end{document}